\documentclass[article,amsmath,11pt,amssymb,floatfix,superscriptaddress,
showpagenumber,nofootinbib]{revtex4}
\usepackage{graphicx}
\usepackage{epsfig}
\usepackage{bm}
\usepackage{amsfonts}
\usepackage{color}
\input epsf

\begin{document}

\title{Reconstruction schemes of scalar field models for the Power Law Entropy Corrected Holographic Dark Energy model with Ricci scalar cut-off }

\author{Antonio Pasqua }
\email{toto.pasqua@gmail.com} 
\affiliation{Department of Physics, University of Trieste, Trieste, Italy.}

\author{Surajit Chattopadhyay}
\email{surajitchatto@outlook.com;~schattopadhyay1@kol.amity.edu}
\affiliation{Department of Mathematics, Amity University, Kolkata, Major Arterial Road, Action Area II, Rajarhat, New Town, Kolkata 700135,
India.}

\author{Irina Radinschi}
\email{radinschi@yahoo.com}
\affiliation{Department of Physics, "Gheorghe Asachi" Technical University, 700050 Iasi, Romania}

\author{Azzah Aziz Alshehri}
\affiliation{University of Hafr Al Batin, Al Jamiah, Hafar Al Batin 39524, KSA}

\author{Abdel Nasser Tawfik}
\email{a.tawfik@fue.edu.eg}
\affiliation{Future University in Egypt (FUE), Fifth Settlement, End of 90th Street, 11835 New Cairo, Egypt}

\begin{abstract}
In this work, we examine the cosmological characteristics of the Power Law Entropy Corrected Holographic Dark Energy (PLECHDE) model with infrared (IR) cut-off, which is determined by the curvature parameter $k$, the time derivative of $H$, and the average radius of the Ricci scalar curvature $R$, which varies with the Hubble parameter $H$ squared.  We obtain the deceleration parameter $q$ and the Equation of State (EoS) parameter of Dark Energy (DE) $\omega_D$. Additionally, we derive the Hubble parameter $H$ and the scale factor $a$ expressions as functions of the cosmic time $t$. Additionally, we examine the limiting scenario that pertains to a flat Dark Dominated Universe. Furthermore, we establish a correspondence between the DE model considered and  some scalar fields, in particular  the Generalized Chaplygin Gas, the Modified Chaplygin Gas, the Modified Variable Chaplygin Gas, the New Modified Chaplygin Gas, the Viscous Generalized Chaplygin Gas, the Dirac-Born-Infeld, the Yang-Mills, and the Non Linear Electrodynamics scalar field models.
\end{abstract}


\date{\today}

\maketitle

\tableofcontents

\section{Introduction}
The current Universe is passing through a phase of an accelerated expansion which was independently reported by Reiss  $\emph{et al.}$ \cite{1b} and Perlmutter $\emph{et al.}$  \cite{1a} in the late $90$'s. Recently, the cosmological and astrophysical data obtained with Supernovae Ia (SNeIa), Large Scale Structure (LSS), the Cosmic Microwave Background (CMB) radiation anisotropies and X-ray experiments have given evidences and supported the late time accelerated expansion of the Universe \cite{1b,1a,1,1c,1d,1h,cmb2,cmb3,planck,sds1,sds2,xray}.
Thus, scientists began to investigate the possible causes of this accelerated expansion and found the suitable models in this context \cite{Tawfik:2019dda,Tawfik:2017ngn}. The models proposed so far to explain the observed accelerated expansion of the Universe are classified in three main cases
\begin{enumerate}
\item The cosmological constant $\Lambda$,
\item Dark Energy (DE) models, and
\item  Modified Gravity theories.
\end{enumerate}
The cosmological constant $\Lambda$ is the simplest candidate of DE, having constant Equation of State (EoS) parameter $\omega = -1$. For the sake on completeness, the equation of state of ordinary matter is recently reported \cite{Tawfik:2019jsa,Ezzelarab:2015tya,NasserTawfik:2012jpa}. In order to explain the observational evidences of accelerated expansion of the Universe, the $\Lambda$ represents the earliest and simplest candidate. Although it is having EoS parameter $\omega =p/\rho= -1$, which is consistent with the observation but it cannot give the time evolution of the EoS parameter. It suffers from two main difficulties \cite{copeland-2006}
\begin{enumerate}
\item the fine-tuning: It asks why the vacuum energy density is so small (about $10^{123}$ lower than what we observe) and
\item the cosmic coincidence problems: It says why vacuum energy and DM are nearly equal today.
\end{enumerate}
It is conjectured that vacuum energy and dark matter (DM) have evolved independently from different mass scales \cite{Chattopadhyay2020,Chattopadhyay2016}. Till date many attempts have been made to find a solution to the coincidence problem \cite{delcampo,delcampoa,delcampob,delcampoc,delcampod,delcampoe,delcampof}.

The second class of models proposed to describe the accelerated expansions of the Universe are DE models. DE is an exotic matter characterized by negative pressure and is thought to be responsible for driving this acceleration. The cosmic acceleration evidence suggests that if Einstein's theory of General Relativity is valid on cosmological scales, the Universe must be dominated by a mysterious and unknown kind of missing component with some peculiar features, for example it must not be clustered on large length scales and its pressure $p$ must be negative enough to be able to drive the accelerated expansion. To study cosmic acceleration, we can observe that it can be described by a perfect fluid with pressure $p$ and energy density $\rho$ satisfy the relation $\rho + 3p < 0$.  This kind of fluid with negative pressure is dubbed as DE. It has EoS parameter $\omega$ (defined as $ p/\rho$) must obey the condition $\omega <-1/3$. Unlike EoS parameter of $\Lambda$, it is a difficult task to constrain its exact value from an observational point of view. The DE model has quintessence behaviour if $\omega > -1$and the model has phantom behaviour if $\omega <-1$. If $\omega$ transits from quintessence to phantom then it is quantum. The largest amount of the total cosmic energy density $\rho_{tot}$ is contained in the dark sectors, i.e. Dark Energy (DE) and Dark Matter (DM).  From the references \cite{reff3, reff4}, currently the DE is 68.3$\%$ of the $\rho_{tot}$. DM contributes about 26.8$\%$ of the $\rho_{tot}$ of the present Universe. The ordinary baryonic matter are able to observe with the scientific instruments and suggests that it contributes only 4.9$\%$ of $\rho_{tot}$. Moreover, radiation contributes to the total cosmic energy density in a practically negligible way. In the literature, other candidates of DE have been proposed, and those having time-varying EoS parameters. Some of them include tachyon, quintessence,  k-essence, quantum, Chaplygin gas, Agegraphic DE (ADE)  and phantom \cite{dil1,dil2,dil2-1,kess1,kess3,kess4,kess2-2,quint1,quint3,quint5,quint6,quint7,tac3,tac1-1,tac1-2,tac1-3,tac1-4,tac2-1,tac2-3,tac2-4,pha1,pha3,pha4,pha5,pha6,pha7,qui1,qui2,qui3,qui4,qui5,qui6,qui8,qui10,qui12,cgas1,cgas2,cgas3,ade1,ade2,2,2a}.

In literature, one of the most studied DE candidates is the Holographic DE (HDE) \cite{3, reff6, 16c, reff8, reff9, ricci26, reff11}. HDE is based on the Holographic Principle (HP) \cite{3,3b,4,5,5a,5b,6}. The HDE model fits the cosmological data obtained from CMB radiation anisotropies and SNeIa \cite{16,16a,16b,16c,16d,16e}. It was shown by Cohen  $\emph{et al.}$   \cite{7} that in Quantum Field Theory (QFT), the ultraviolet (UV) cut-off, indicated with $\Lambda_{UV}$, should be related to the IR cut-off $L$ due to limit set by forming a black hole. If the vacuum energy density caused by UV cut-off is given by $\rho_D = \Lambda^4_{UV}$, then the total energy of size $L$ should not be greater than the mass of the system-size black hole, i.e.
\begin{eqnarray}
E_D \leq E_{BH}, \label{1old},
\end{eqnarray}
which yields
\begin{eqnarray}
L^3 \rho_D \leq M_p^2 L, \label{1},
\end{eqnarray}
where the quantity $M_p = \left( 8\pi G  \right)^{-1/2} \approx 10^{18}~$GeV represents the reduced Planck mass (with $G$ being the Newton's gravitational constant). If the largest possible cut-off $L$ is the one which saturate the inequality given in Eq. (\ref{1}), we obtain the energy density of HDE $\rho_D$ as follows.
\begin{eqnarray}
\rho_D = 3c^2 M_p^2 L^{-2}, \label{2}
\end{eqnarray}
where $c$ represents a dimensionless numerical constant which value can be evinced by observational data. For  a flat Universe $c=0.818_{-0.097}^{+0.113}$  and in the case of a
non-flat Universe, we have  $c=0.815_{-0.139}^{+0.179}$ \cite{n2primo,n2secondo}. According to a recent work made by Guberina  $\emph{et al.}$  \cite{8}, the HDE model based on the entropy bound can be derived in an alternative way. In the black hole thermodynamics \cite{9,9a}, a maximum entropy in a box with size $L$, known as Bekenstein--Hawking entropy bound, exists and it is given by $S_{BH} \approx M_p^2 L^2$;  moreover, it scales as the area of the box (i.e. $A \approx L^2$) rather than its volume (i.e. $V \approx L^3$). Moreover, for a macroscopic system with self-gravitation effects which can not be ignored, the Bekenstein entropy bound $S_B$ can be obtained by multiplying the energy  and the linear size $L$ of the system $E \approx \rho_DL^3$. If we require that the Bekenstein entropy bound is smaller than the Bekenstein--Hawking entropy (i.e. $S_B \leq S_{BH}$, which implies $EL \leq M_p^2 L^2$), it is possible to obtain the same result obtained from energy bound argument, i.e.
\begin{eqnarray}
 \rho_D \leq M_p^2 L^{-2}.
 \end{eqnarray}
The HDE model has been studied widely and investigated in literature in many ways till now. Authors obtained that EoS parameter of HDE model can be significantly reconstruct in the low redshift limit. Chen  $\emph{et al.}$   \cite{10} used the HDE model in order to drive inflation in the early evolutionary phases of the Universe. Jamil  $\emph{et al.}$   \cite{11} studied the EoS parameter $\omega_D$ of the HDE model considering a time-varying Newton's gravitational constant, i.e. $G \equiv G\left( t \right)$.

The HDE models studied by modifying infrared (IR) cut-off and with different IR cut-offs also \cite{12,12a,12c,12e,12f,12g,13,13c}, for example by taking correspondence of IR cut-off with the particle horizon, the future event horizon and the Hubble horizon. Recently, the correspondence between HDE models and other scalar field models have been proposed \cite{14,14a,14b}. The late time acceleration of the Universe has also been investigated by the concept of modified gravity \cite{nojo,nojo2}. Modified Gravity is predicted by string/M theory. It gives the natural gravitational alternative to the idea of the presence of exotic components. The explanation of the phantom, non-phantom and quantum phases of the Universe can be well described using modified gravity without the necessity of the introduction of a negative kinetic term in DE models.

Many dynamical features of HDE model have been studied in the flat/non-flat Friedmann-Lemaitre-Robertson-Walker (FLRW) background, for example the cosmic coincidence problem, the quantum behavior, the phantom crossing at the present time, the EoS parameter $\omega_D$ and the deceleration parameter $q$.

The definition of HDE can be modified due to the power-law corrections to entropy which appear in dealing with the entanglement of quantum fields in and out the horizon \citep{pl1}.
The entanglement entropy of the ground state obeys the ``Hawking area law''. Only the excited state gives a contribution to the correction and more excitation produces more deviations from the area law \citep{pl3,pl4}.
The power-law corrected entropy has the following form \citep{pl1,pl2}:
\begin{eqnarray}
    S\left( A \right)_{pl} =c_0 \left( \frac{A}{a_1^2}  \right)\left[ 1+c_1f\left( A \right)  \right],\label{powerlawentropyold}
\end{eqnarray}
where the term $f(A)$ is given by
\begin{eqnarray}
 f\left( A \right)  = \left( \frac{A}{a_1^2}  \right)^{-\nu}, \label{powerlawentropyold2}
\end{eqnarray}
i.e., $f(A)$ has a power-law dependence from the area $A$. The two quantities $c_0$ and $c_1$ are two constant parameters of the order of unity. The UV cut-off at the horizon is denoted by $a_1$ and $\nu$ indicates an exponent which depends on the amount of mixing of ground and excited states. $A=4\pi R_h^2$ indicates the area of the horizon, where $R_h$ is the black-hole event horizon. The contribution of the term $f \left(A\right)$ is negligible and the mixed state entanglement entropy asymptotically approaches the ground state (Bekenstein--Hawking) entropy for large horizon area (i.e., $A>>a_1^2$) \cite{Yassin:2020mjf,Tawfik:2016tfe,Tawfik:2016wfm,Tawfik:2015fda}.

Another useful form of the power-law corrected expression of the entropy is given by the following relation:
\begin{eqnarray}
    S\left( A \right)_{pl}=\frac{A}{4G}\left(1-K_{\alpha}A^{1-\alpha/2}   \right), \label{powerlawentropy}
\end{eqnarray}
where $\alpha$ is a dimensionless constant  and the term $K_{\alpha}$ is a constant parameter whose expression is given by
\begin{eqnarray}
K_{\alpha}=\frac{\alpha \left( 4\pi \right)^{\alpha/2-1}}{\left(4-\alpha \right)r^{2-\alpha}_c}. \label{murano1}
\end{eqnarray}
The quantity $r_c$ represents the crossover scale.
When the wave function of the field is chosen to be a superposition of ground and exited states \citep{pl3}, the second term in Eq. (\ref{powerlawentropy}) can be regarded as a power-law correction to the area law resulting from entanglement  \citep{pl3}. The correction term is also more significant for higher excitation \citep{pl3}. It is important to notice that the correction term falls off rapidly with the area $A$ and hence in the semi-classical limit (which means large values of $A$) the area law is recovered. Then, for large black holes the correction term falls off rapidly and the area law is recovered, whereas for small black holes the correction becomes more significant.
This can be interpreted as follows. For a large area, i.e., at low energies, it is difficult to excite the modes and hence the ground state modes contribute to most of the entanglement entropy. However, for a small horizon area, a large number of field modes can be excited and contribute significantly to the correction, causing large deviation from the area law.

Inspired by the power--law corrected entropy relation given in Eq. (\ref{powerlawentropy}), the energy density of the so-called Power-Law Entropy Corrected Holographic Dark Energy (PLECHDE) model can be easily obtained as follows.
\begin{eqnarray}
    \rho_{Dpl}&=&3c^2 M_p^2L^{-2}-\delta M_p^2L^{-\gamma} \nonumber \\
    &=& 3M_p^2L^{-2}\left[ c^2 -\left(\frac{\delta}{3}\right)L^{-\gamma +2}   \right]. \label{newrhoold}
\end{eqnarray}
In the limit case corresponding to $\delta = 0$, the expression of the energy density of DE $\rho_D$ of the PLECHDE given in Eq. (\ref{newrhoold}) yields the well-known holographic energy density, i.e., $\rho_D=3 c^2 M_p^2L^{-2}$. From a mathematical point of view, the HDE model can be also recovered in the limiting case of $\gamma \rightarrow \infty$.

The importance of the corrected term in various regions depends on the value assumed by $\gamma$. The two terms of Eq. (\ref{newrhoold}) can be combined when $\gamma = 2$  and can recover the ordinary HDE density. So, it is possible to study the cases with $\gamma > 2$ and $\gamma < 2$ separately. In the first case, i.e., $\gamma > 2$, the corrected term can be comparable to the first term only when $L$ is very small. Indeed, it was argued that $\gamma$ should be within the range $2 < \gamma < 4$ \citep{pl3}. However, the satisfaction of the generalized second law of thermodynamics for the Universe with the power-law corrected entropy given in Eq. (\ref{powerlawentropy}) implies that the case $\gamma < 2$ should be rejected \citep{pl2}.

In this paper, we consider the IR cut-off $L$ of the system proportional to the average radius of Ricci scalar curvature $R$, i.e., $L\propto R^{-1/2}$, so that we have $\rho_D \propto R$. We remember that the Ricci scalar curvature $R$ can be written as follows
\begin{eqnarray}
    R=6\left[ \dot{H}+2H^2+\frac{k}{a\left(t\right)^2}\right],\label{4}
\end{eqnarray}
where $H=\dot{a}\left(t\right)/a\left( t \right)$ is the Hubble parameter, $\dot{H}$ is the first derivative of $H$ with respect to the cosmic time $t$, $a\left(t\right)$ is a dimensionless scale factor, which describes how the Universe evolves, and $k$ is a constant with dimension of $length^{-2}$ known as curvature parameter which contains the information about the curvature of the spatial part of the metric describing the Universe, which will be introduced later on. $k$ takes the values $-1,\, 0,\, +1$ which yield, respectively, an open, a flat or a closed FRW Universe.

Ricci DE (RDE) models have been widely studied in recent years  \cite{ricci1,ricci2,ricci4,ricci5,ricci6,ricci7,ricci8,ricci9,ricci10,ricci11,ricci12,ricci13,ricci14,
ricci16,ricci17,ricci19,ricci22,ricci23,ricci24,ricci26,ricci27,mio1,mio2}.
Substituting $L$ with $R^{-1/2}$ in Eq.(\ref{4}) and considering $M_p^2=1$, we can write the energy density of the Ricci PLECHDE (R-PLECHDE) model $\rho_{Dpl}$ as follows.
\begin{eqnarray}
    \rho_{Dpl}=3c^2 R-\delta R^{\gamma /2}.\label{murano2}
\end{eqnarray}
In the following Sections we will derive and study some important cosmological quantities considering the energy density of DE defined in Eqs. (\ref{murano2}).

This paper is organized as follows. In Section 2, we describe the physical contest we are working in and we derive the EoS parameter $\omega_D$, the deceleration parameter $q$ and $\Omega_D'$ for our models in a non-flat Universe. In Section 3, we establish a correspondence between our model and some scalar fields, in particular  the Generalized Chaplygin Gas (GCG), the Modified Chaplygin Gas (MCG), the Modified Variable Chaplygin Gas (MVCG), the New Modified Chaplygin Gas (NMCG), the Viscous Generalized Chaplygin Gas (VGCG), the Dirac-Born-Infeld (DBI),  the Yang-Mills (YM) and the Non Linear Electro-Dynamics scalar (NLED) field models. Section 4 is devoted to the Conclusions of this paper.

\section{NON INTERACTING MODEL IN A NON-FLAT UNIVERSE}
Observational and cosmological evidences suggest that our Universe is not perfectly flat but it has a small positive curvature, which implies a closed Universe. The tendency of a closed Universe is shown in cosmological (in particular CMB) experiments \citep{closed2,closed4,closed5}. Moreover, the measurements of the cubic correction to the luminosity-distance relation of Supernova measurements reveal a closed Universe \citep{cubic1,cubic2}. For the above reasons, we prefer to consider a non-flat Universe.

Within the framework of the standard Friedmann-Robertson-Walker (FRW) cosmology, the line element for a non-flat Universe is given by the following relation:
\begin{eqnarray}
    ds^2=-dt^2+a^2\left(t\right)\left[\frac{dr^2}{1-kr^2} +r^2 \left(d\theta ^2 + \sin^2 \theta d\varphi ^2\right) \right],\label{6}
\end{eqnarray}
where $t$ represents the cosmic time, $r$ is the radial component of the metric, $\theta$ and  $\varphi$ are the two angular coordinates of the metric.   $\theta$ and $\varphi$ are the usual azimuthal and polar angles, with the constraints given by $0\leq \theta \leq \pi$ and $0\leq \varphi \leq 2\pi$. 
The corresponding Friedmann equation for a non-flat Universe has the following expression:
\begin{eqnarray}
    H^2+\frac{k}{a^2}=\frac{1}{3 }\left( \rho _D + \rho _{m}\right),\label{7}
\end{eqnarray}
where the terms $\rho _D$ and $\rho _{m}$ are, respectively, the energy densities of DE and DM. We also define the fractional energy densities for matter, curvature and DE, respectively, as
\begin{eqnarray}
    \Omega _m &=& \frac{\rho_m}{\rho _{cr}} = \frac{\rho _m}{3 H^2},\label{8}\\
    \Omega _k &=& \frac{\rho_k}{\rho _{cr}} = \frac{k}{H^2a^2},\label{9}\\
    \Omega _{Dpl} &=& \frac{\rho_{Dpl}}{\rho _{cr}} = \frac{\rho _{Dpl}}{3 H^2},\label{10}
\end{eqnarray}
where $\rho_{cr} = 3 H^2$ represents the critical energy density, which represents the energy density necessary for flatness. The fractional energy density of curvature $\Omega _k$ gives information about the contribution from the spatial curvature to the total density. Recent observations support a closed Universe with a small positive curvature $\Omega _k \cong 0.02$ \citep{spergel}.

Dividing Eq. (\ref{7}) by $H^2$ and using Eqs. (\ref{8}), (\ref{9}) and (\ref{10}), it is possible to write the Friedmann equation given in Eq. (\ref{7}) as
\begin{eqnarray}
1 + \Omega _k = \Omega _{m} + \Omega _{Dpl}. \label{11pl}
\end{eqnarray}
Eq. (\ref{11pl})  has the main property that it  relates all the fractional energy densities considered in this paper. If we want to respect the Bianchi identity or local energy-momentum conservation law, i.e., if we want to satisfy the relation $\nabla_{\mu}T^{\mu \nu}=0$, then we must have that the total energy density $\rho_{tot}$ must satisfy the following continuity equation:
\begin{eqnarray}
    \dot{\rho}_{tot}+3H\left( p_{tot} +\rho_{tot}\right)=0, \label{12},
\end{eqnarray}
where the terms $\rho_{tot}$ and $p_{tot}$ are the total energy density and the total pressure, respectively, and they are defined as follows.
\begin{eqnarray}
    \rho_{tot}&=&\rho_m + \rho_{Dpl},\\
    p_{tot}&=&p_{Dpl}.
\end{eqnarray}
We must emphasize here that we are considering pressureless DM, i.e., we have $p_m=0$, therefore we have that the total pressure coincides with the DE pressure.

We have that Eq. (\ref{12}) can be also rewritten as follows.
\begin{eqnarray}
    \dot{\rho}_{tot}+3H\left( 1+\omega_{tot} \right)\rho_{tot}=0\label{12altro},
\end{eqnarray}
where $\omega_{tot}= p_{tot}/\rho_{tot}$ represents the Equation of State (EoS) parameter.

Since the two energy densities $\rho_D$ and $\rho_m$ are conserved separately, the conservation equations for DM and DE take the forms
\begin{eqnarray}
    \dot{\rho}_m+3H\rho_m &=& 0, \label{13}\\
    \dot{\rho}_{Dpl}+3H\rho_{Dpl}\left( 1 + \omega_{Dpl}\right)&=&0.\label{14pl}
\end{eqnarray}
We now consider the presence of interaction between the two dark sectors.

By assuming an interaction between DM and DE, the conservation equations for DM and DE take the following forms
\begin{eqnarray}
    \dot{\rho}_m+3H\rho_m &=& Q, \label{13int}\\
        \dot{\rho}_{Dpl}+3H\rho_{Dpl}\left( 1 + \omega_{Dpl}\right)&=&-Q,\label{14plint}
\end{eqnarray}
where the quantity $Q$ indicates an interaction term which can be, in general, a function of cosmological quantities like the Hubble parameter $H$, energy densities for DE and DM $\rho_D$ and $\rho_m$ and the deceleration parameter $q$, i.e., we have that $Q\left(H,\rho_m,\rho_D,q\right)$. In this paper, we have chosen to consider three different expressions for $Q$, i.e.,
\begin{eqnarray}
    Q_1 &=& 3b^2H\left(\rho _m + \rho _D\right),\label{15}\\
    Q_1 &=& 3b^2H\rho _m ,\label{15-1}\\
    Q_1 &=& 3b^2H\rho _D,\label{15-2}
\end{eqnarray}
with $b^2$ being a coupling parameter between DM and DE \citep{q1,q3,q4,q5,q6,q9,q10,q11}. We must anyway underline that more general interaction terms can be used \citep{q12}. Since the nature of DM and DE remains unknown, different Lagrangian equations have been proposed to generate this interaction term. Positive values of $b^2$ indicate a transition from DE to DM and vice versa for negative values of $b^2$. Sometimes $b^2$ is taken in the range [0,1] \citep{q13}. The case with $b^2 = 0$ represents the non-interacting FRW model while $b^2 = 1$ yields the complete transfer of energy from DE to matter. Recently, it is reported that this interaction is observed in the Abell cluster A586 showing a transition of DE into DM and vice versa \citep{q14,q15}. However, we must underline here that the strength of this interaction is not clearly identified \citep{q16}. Observations of both CMB radiation and clusters of galaxies clearly show that the coupling parameter  must be in the range $0<b^2 < 0.025$ \citep{q17,q18}. A negative value of $b^2$ is avoided since it  leads to a violation of thermodynamic laws. Further high-resolution N-body simulations have demonstrated that the structural properties of highly nonlinear cosmic structures, as  their average concentration at a given mass, could be significantly modified in the presence of an interaction between DE and DM \cite{stabilo1}. The strength of the coupling parameter can, in fact, significantly modify the cosmic history by modifying the clustering properties of matter since the growth of DM density perturbations is much more sensitive to the interaction \cite{stabilo2,stabilo3}. The best way to motivate a suitable form of $Q$ should be from a consistent theory of quantum gravity or through a reconstruction scheme using the SNeIa data \cite{stabilo4,stabilo5}.

We now want to derive the expression of the EoS parameter of DE $\omega_D$ for the models we are studying in the case the IR cut-off of the system is given by the Ricci scalar curvature.

Using the Friedmann equation given in Eq. (\ref{7}) in the general expression of the Ricci scalar $R$ given in Eq. (\ref{4}), we obtain that the Ricci scalar $R$ can be also written as
\begin{eqnarray}
    R=6\left(  \dot{H} + H^2 + \frac{\rho_m+\rho_D}{3} \right). \label{17}
\end{eqnarray}
We now want to obtain an expression for the term $ \dot{H} + H^2$ as function of the EoS parameter of DE $\omega_D$ in order we are able to relate the EoS parameter $\omega_D$ and $R$.

Differentiating the Friedmann equation given in Eq. (\ref{7}) with respect to the cosmic time $t$, we obtain
\begin{eqnarray}
   2H \dot{H} -2H\left(\frac{k}{a^2}\right)=\frac{1}{3}\left( \dot{\rho}_D + \dot{\rho}_m  \right). \label{dotHprime}
\end{eqnarray}
When dividing Eq. (\ref{dotHprime}) by $2H$, we get
\begin{eqnarray}
    \dot{H}=\frac{k}{a^2}-\frac{1}{6H}\left( \dot{\rho}_D + \dot{\rho}_m  \right). \label{dotH}
\end{eqnarray}
By inserting in Eq. (\ref{dotH}) the expressions of the time derivatives of the energy densities of DM and DE $\dot{\rho}_m$ and $\dot{\rho}_{Dpl}$  obtained from the continuity equations for DM and DE given in Eqs. (\ref{13}) and (\ref{14pl}), we can rewrite $\dot{H}$ as
\begin{eqnarray}
\dot{H}    = \frac{k}{a^2} - \frac{1}{2} \left[ \rho_m + \left( 1+\omega_D \right)\rho_D  \right]. \label{18}
\end{eqnarray}
Adding Eqs. (\ref{7}) and (\ref{18}), we obtain the following relation for $\dot{H} + H^2$
\begin{eqnarray}
\dot{H}+H^2=-\frac{1}{6}\left(\rho_m+\rho_D\right)-\frac{\omega_D \rho_D }{2}.   \label{jamil20}
\end{eqnarray}
Therefore, inserting the result of Eq. (\ref{jamil20}) into Eq. (\ref{17}), after some algebraic calculations, we obtain that the Ricci scalar $R$ can be written as follows.
\begin{eqnarray}
R &=& \rho_m + \rho_D-3\rho_D\omega_D \nonumber \\
&=& \rho_m +\rho_D \left( 1- 3 \omega_D  \right).   \label{jamil22}
\end{eqnarray}
From the result of Eq. (\ref{jamil22}), we can easily derive that the expression of the EoS parameter of the PLECHDE model $\omega_{Dpl}$ is given by the following relation:
\begin{eqnarray}
\omega_{Dpl}&=&-\frac{R}{3\rho_{Dpl}}+\frac{\Omega_{Dpl}+\Omega_{m}}{3\Omega_{Dpl}} = -\frac{R}{3\rho_{Dpl}}+\frac{1+\Omega_{k}}{3\Omega_{Dpl}} \nonumber \\
&=& -\frac{R}{3\rho_{Dpl}} + \frac{1+u_{pl}}{3}= \frac{1}{3} \left(1+u_{pl}-\frac{R}{\rho_{Dpl}}  \right),\label{21}
\end{eqnarray}
where we used the fact that $\frac{\rho_D+\rho_{m}}{3\rho_D}=\frac{\Omega_D+\Omega_{m}}{3\Omega_D} $ while the parameter $u_{pl}$  is defined as
\begin{eqnarray}
 u_{pl}&=&\frac{\rho_{m}}{\rho_{Dpl}},\label{miami1}
\end{eqnarray}
which is equivalent to the following expression:
\begin{eqnarray}
 u_{pl}&=&\frac{\Omega_{m}}{\Omega_{Dpl}},\label{miami11}
\end{eqnarray}
Moreover, using the expressions of the fractional energy density given in Eq. (\ref{11pl}), we obtain
\begin{eqnarray}
 u_{pl}&=&\frac{1+\Omega_{k}}{\Omega_{Dpl}}-1,\label{miami111}
\end{eqnarray}
Substituting into Eq. (\ref{21}) the expression of the energy density $\rho_{Dpl}$ given in Eq. (\ref{murano2}) and using the relation between the fractional energy densities given in Eq. (\ref{11pl}) along with the definition of $u_{pl}$, we obtain the following expression of the EoS parameter for the R-PLECHDE model $\omega_{Dpl}$
\begin{eqnarray}
\omega_{Dpl} &=& -\frac{1}{3} \left(  \frac{1}{3c^2 -\delta R^{\gamma /2-1}}-  \frac{1+\Omega_k}{\Omega_{Dpl}}\right)\nonumber \\
&=&  -\frac{1}{3}\left[ \frac{1}{3c^2 -\delta R^{\gamma /2-1}}- \left(1+u_{pl}\right)\right]. \label{schirinzi}
\end{eqnarray}

For completeness, we also derive the expression of the deceleration parameter $q$. The deceleration parameter $q$ is generally defined as
\begin{eqnarray}
    q&=&-\frac{\ddot{a}a}{\dot{a}^2} = -1-\frac{\dot{H}}{H^2}.\label{31}
\end{eqnarray}
$q$, combined with the Hubble parameter $H$ and the dimensionless density parameters, form a set of very useful parameters for the description of the astrophysical observations. The expansion of the Universe becomes accelerated if the term $\ddot{a}$ has a positive value, as recent cosmological observations  suggest;  in this case, $q$ assumes a negative value. Differentiating  the Friedmann given in Eq. (\ref{7}) with respect to the cosmic time $t$ and using the results of Eq. (\ref{11pl})  along with the continuity equations for DM and DE in Eq. (\ref{31}), it is possible to write the expression of the deceleration parameter $q$  as
\begin{eqnarray}
    q=\frac{1}{2}\left(1 + \Omega_k + 3\Omega_D \omega_D  \right).\label{32}
\end{eqnarray}
Substituting in Eq. (\ref{32}) the expression of the EoS parameter of DE $\omega_{Dpl}$ of the R-PLECHDE model given in Eq. (\ref{schirinzi}), we obtain the following expression of $q_{pl}$ for the R-PLECHDE model
\begin{eqnarray}
    q_{pl}=1-\frac{1}{2}\left(\frac{\Omega _D}{3c^2 -\delta R^{\gamma /2-1}}\right)+ \Omega _k.
\end{eqnarray}
We can now derive some important quantities of the R-PLECHDE  model in the limiting case of a flat Dark Dominated Universe, which is recovered when $\delta=0$, $\Omega_{Dpl}=1$, $\Omega_k=\Omega_m=0$ and $u_{pl}$.

The expression of the energy density of DE model reduces to
\begin{eqnarray}
\rho_D=3c^2  R,\label{34old}
\end{eqnarray}
where $R$, in the case corresponding to a flat Dark Dominated Universe, is given by
\begin{eqnarray}
R = 6 \left( \dot{H} + 2H^2  \right), \label{RDD}
\end{eqnarray}
since we have that $k=0$.

Therefore, the expression of $\rho_D$ given in Eq. (\ref{34old}), using the expression of the Ricci scalar $R$ given in Eq. (\ref{RDD}), can be written as
\begin{eqnarray}
\rho_D=18c^2  \left( \dot{H} + 2H^2  \right).\label{34}
\end{eqnarray}
Inserting the expression of $\rho_D$ given into Eq. (\ref{34}) in the Friedmann equation given in Eq. (\ref{7}), we obtain the following first order differential equation for the Hubble parameter $H$
\begin{eqnarray}
\dot{H} + \left(\frac{12c^2 -1 }{6c^2}\right)H^2 =0,\label{murano3}
\end{eqnarray}
whose solution is given by
\begin{eqnarray}
H\left(t \right)=\left(\frac{6c^2}{12c^2-1}\right)\left(\frac{1}{t}\right).\label{35}
\end{eqnarray}
In order to have a well defined expression of the Hubble parameter $H$, we must have that $c^2>1/12$.

Inserting the expression of $H$ derived in Eq. (\ref{35}) into the expression of the Ricci scalar curvature for a flat Dark Dominated Universe obtained in Eq. (\ref{RDD}), we obtain the following expression for the Ricci scalar $R$ as a function of the cosmic time $t$
\begin{eqnarray}
R\left(t \right)=\frac{36c^2}{(12c^2-1)^2}\left(\frac{1}{t^2}\right).\label{36}
\end{eqnarray}
Using the result of Eq. (\ref{36}) in Eq. (\ref{34old}), we obtain the following result for the energy density of DE $\rho_D$ as a function of the cosmic time $t$
\begin{eqnarray}
\rho_D\left(t \right)&=&\frac{108c^4}{(12c^2-1)^2}\left(\frac{1}{t^2}\right) \nonumber \\
&=&3\left(\frac{ 6c^2 }{12c^2-1}\right)^2\left(\frac{1}{t^2}\right).\label{rhoschi}
\end{eqnarray}
At last, the EoS parameter of DE $\omega_D$ and the deceleration parameter $q$ reduce, respectively, to
\begin{eqnarray}
    \omega_D&=&\frac{1}{3}-\frac{1}{9c^2}, \label{LEos}\\
    q&=&1-\frac{1}{6c^2}. \label{Lq}
\end{eqnarray}
From Eq. (\ref{LEos}), we obtain that, in the limiting case of a flat Dark Dominated Universe, the EoS parameter of DE assumes a constant value and, for $c^2<1/12$, we obtain $\omega_D<-1$, i.e., the phantom divide line can be crossed. Since the Ricci scalar diverges at $c^2=1/12$, this value of $\alpha$ can not be taken into account. Furthermore, from Eq. (\ref{Lq}), we obtain that the accelerated phase of the Universe (which is recovered for $q<0$) starts for $c^2 \leq 1/6$, which corresponds to the starting of the quintessence regime $(\omega_D \leq -1/3)$.

We can also observe that Eqs. (\ref{LEos}) and (\ref{Lq}) are related by the following relation:
\begin{eqnarray}
\omega_D = \frac{2q}{3}-\frac{1}{3}.\label{bubi}
\end{eqnarray}
We must also emphasize here that the case we obtained is similar to power-law expansion of scale factor obtained in \citep{granda2008}, in which $a(t)=t^{6c^2/(12c^2-1)}$.

Considering the value of $c^2$ obtained in the paper of Gao \cite{gaoricci}, i.e., $c^2 \approx 0.46$, we obtain the following values for the cosmological quantities we are dealing with 
\begin{eqnarray}
H(t)&\approx& \frac{0.610}{t}, \\
R(t)&\approx& \frac{0.811}{t^2}, \\
\rho_D(t) &\approx & \frac{1.119}{t^2},\\
\omega_D &\approx& 0.092,\\
q&\approx&0.638,\\
a(t)&\approx& t^{0.610}.
\end{eqnarray}
Therefore, we have obtained with the value of $c^2$ derived in ref. \cite{gaoricci} a decelerating Universe because of positive value of the deceleration parameter. We now want to derive the evolutionary form of the fractional energy density of DE. Differentiating the expression of $\Omega_{Dpl}$  given in Eq. (\ref{10}) with respect to the variable $x$, we obtain the following expression for $\Omega_{Dpl}'$
\begin{eqnarray}
\Omega_{Dpl}' &=& \Omega_{Dpl}\left[\frac{\rho_{Dpl}'}{\rho_{Dpl}} - 2\left(\frac{\dot{H}}{H^2}\right)_{pl}\right], \label{evoomegadpl}
\end{eqnarray}
We now derive the expression of $\rho_{Dpl}'$ from the continuity equation for DE given in Eq. (\ref{14pl})
\begin{eqnarray}
\rho_{Dpl}' &=& -3\rho_{Dpl}\left(  1+\omega_{Dpl}\right).
\end{eqnarray}
We can find the expression of $\left(\frac{\dot{H}}{H^2}\right)_{pl}$ from Eq. (\ref{jamil20})
\begin{eqnarray}
\left(\frac{\dot{H}}{H^2}\right)_{pl}&=&-1-\frac{1}{6H^2}\left(\rho_m+\rho_{Dpl}\right)-\frac{\omega_{Dpl} \rho_{Dpl} }{2H^2} \nonumber \\
&=& -1-\frac{\rho_{Dpl}}{6H^2}\left(1+u_{pl}\right)-\frac{\omega_{Dpl} \rho_{Dpl} }{2H^2}.   \label{jamil201pl}
\end{eqnarray}
Using the definition of the fractional energy density of DE defined in Eq. (\ref{10}), we can write Eq. (\ref{jamil201pl})  as follows
\begin{eqnarray}
\left(\frac{\dot{H}}{H^2}\right)_{pl} &=& -5 -3\omega_{Dpl}\left(1+\Omega_{Dpl}  \right) - \Omega_{Dpl} \left( 1+u_{pl}\right). \label{cozlowa1}
\end{eqnarray}
Therefore, using the expression of $\rho_{Dpl}'$ obtained in Eqs. (\ref{}) and (\ref{}) along with the results of Eq. (\ref{cozlowa1}),  we obtain the following final expression for $\Omega_{Dpl}'$:
\begin{eqnarray}
\Omega_{Dpl}' &=& \Omega_{Dpl}\left[7+3\omega_{Dpl}\left(1+2\Omega_{Dpl} \right)+ 2 \Omega_{Dpl} \left( 1+u_{pl}\right)  \right]. \label{}
\end{eqnarray}
We now derive the evolutionary form of the parameter $u_{pl}$. Differentiating the expression of $u_{pl}$ given in Eq. (\ref{miami1})  with respect to the cosmic time $t$, we derive that
\begin{eqnarray}
\dot{u}_{pl} &=& \frac{\dot{\rho}_m}{\rho_{Dpl}} - \frac{\rho_m \dot{\rho}_{Dpl}}{\rho_{Dpl}^2} = \frac{\dot{\rho}_m}{\rho_{Dpl}} - u_{pl}\left(\frac{\dot{\rho}_{Dpl}}{\rho_{Dpl}}\right). \label{meniadotupl}
\end{eqnarray}
By using the continuity equations for DE and DM given in Eqs. (\ref{13}) and (\ref{14pl}), we derive, after some calculations, that the time evolution of $u_{pl}$ is governed by
\begin{eqnarray}
\dot{u}_{pl}&=& 3H u_{pl} \omega_{Dpl},
\end{eqnarray}
which leads to the following expression for $u_{pl}'$:
\begin{eqnarray}
u_{pl}'&=& 3u_{pl} \omega_{Dpl}.
\end{eqnarray}
We now consider the interacting case. We still consider the general expression given by
\begin{eqnarray}
\Omega_{Dpl}' &=& \Omega_{Dpl}\left[\frac{\rho_{Dpl}'}{\rho_{Dpl}} - 2\left(\frac{\dot{H}}{H^2}\right)_{pl}\right]. \label{evoomegadintlog}
\end{eqnarray}
From the continuity equation for DE given in Eq. (\ref{14plint}), we obtain the following expressions for $\rho'_{Dpl}$
\begin{eqnarray}
\rho'_{Dpl} &=& -3\rho_{Dpl}\left( 1 + \omega_{Dlog}\right) -\frac{Q}{H}.\label{natacoz1}
\end{eqnarray}
Inserting in Eq. (\ref{evoomegadintpl}) the expression of $\rho'_{Dpl}$  obtained in Eq. (\ref{natacoz1})  along with the expression of $\left(\frac{\dot{H}}{H^2}\right)_{pl}$, we can write $\Omega_{Dpl}'$
\begin{eqnarray}
\Omega_{Dpl}' &=& \Omega_{Dpl}\left[7+3\omega_{Dpl}\left(1+2\Omega_{Dpl} \right)+ 2 \Omega_{Dpl} \left( 1+u_{pl}\right) -\frac{Q}{H\rho_{Dpl}} \right]. \label{}
\end{eqnarray}
Using the expressions of $Q_1$, $Q_2$ and $Q_3$ we have defined in Eqs. (\ref{15}), (\ref{15-1}) and (\ref{15-2}), we finally obtain the following relations
\begin{eqnarray}
\Omega_{Dpl1}' &=& \Omega_{Dpl}\left[7+3\omega_{Dpl}\left(1+2\Omega_{Dpl} \right)+ 2 \Omega_{Dpl} \left( 1+u_{pl}\right) -3b^2\left( 1+u_{pl}  \right) \right], \label{}\\
\Omega_{Dpl2}' &=& \Omega_{Dpl}\left[7+3\omega_{Dpl}\left(1+2\Omega_{Dpl} \right)+ 2 \Omega_{Dpl} \left( 1+u_{pl}\right) -3b^2u_{pl}\right], \label{}\\
\Omega_{Dpl3}' &=& \Omega_{Dpl}\left[7+3\omega_{Dpl}\left(1+2\Omega_{Dpl} \right)+ 2 \Omega_{Dpl} \left( 1+u_{pl}\right) -3b^2 \right]. \label{}
\end{eqnarray}
When $b^2=0$, we recover the same results of the non-interacting case.

Following the same procedure of the non-interacting case, we can also derive that, for the interacting case, the evolutionary form of $u_{pl}$ is governed by the following law:
\begin{eqnarray}
u_{pl}'&=& 3u_{pl} \omega_{Dpl} + \frac{Q\left(1+u_{pl}\right)}{H\rho_{Dpl}}.
\end{eqnarray}
Using the definitions of $Q_1$, $Q_2$ and $Q_3$ we have chosen in Eqs. (\ref{15}), (\ref{15-1}) and (\ref{15-2}),  we derive that  $u'_{pl}$ is given by the following relations:
\begin{eqnarray}
u_{pl1}'&=&  3u_{pl} \omega_{Dpl} +3b^2\left(1+u_{pl}\right)^2, \\
u_{pl2}'&=&  3u_{pl} \omega_{Dpl} +3b^2u_{pl}\left(1+u_{pl}\right), \\
u_{pl3}'&=&  3u_{pl} \omega_{Dpl} +3b^2\left(1+u_{pl}\right).
\end{eqnarray}
We obtained now all the relevant cosmological quantities for the R-PLECHDE model for both cases corresponding to interacting and non-interacting Dark Sectors. Therefore, we can now start to study the correspondence between the R-PLECHDE model and the scalar fields model we are considering.

\section{CORRESPONDENCE BETWEEN THE R-PLECHDE MODEL  AND SCALAR FIELDS}

In this Section, we want to establish a correspondence between the R-PLECHDE model and the following scalar fields models: the Generalized Chaplygin Gas (GCG), the Modified Chaplygin Gas (MCG), the Modified Variable Chaplygin Gas (MVCG), the New Modified Chaplygin Gas (NMCG), the Viscous Generalized Chaplygin Gas (VGCG), the Dirac-Born-Infeld (DBI),  the Yang-Mills (YM) and the Non Linear Electro-Dynamics (NLED) models \cite{Tawfik:2017ngn,Tawfik:2016dvd}. We take this decision since it is widely accepted that scalar field models are an effective description of the DE theory.  For this reason, we start comparing the energy density of the DE model considered in this paper  with the energy density of corresponding scalar field model and later on we equate the EoS parameter of scalar field models we have chosen to study with the EoS parameter of the DE  model we studied.

\subsection{Generalized Chaplygin Gas (GCG) Model}

We start making a correspondence between the R-PLECHE model and the first scalar field model considered, i.e., the Generalized Chaplygin Gas (GCG) model. In a recent work, Kamenshchik  $\emph{et al.}$  \cite{gcg1} considered an homogeneous model known as Chaplygin Gas (CG) model which is  based on a single fluid obeying the Equation of State EoS $p=-\frac{A_0}{\rho}$, where $p$ and $\rho$ represent, respectively, the pressure and the energy density of the fluid while the quantity $A_0$ is a positive constant parameter. Authors proposed a generalization known as Generalized Chaplygin Gas (GCG) model.

The GCG model has the property that it can interpolate the evolution of Universe from the dust to the accelerated phase, therefore it can fit the observational cosmological data \cite{gcg6}.

The GCG EoS is defined as follows \cite{gcg8,gcg9,gcg16,gcg17}.
\begin{eqnarray}
    p_D=-\frac{D}{\rho_D^{\theta}}, \label{gcg1}
\end{eqnarray}
where $D$ and $\theta$ are two free constant parameters, with $D$ positive defined and $\theta$  in the range $0<\theta<1$. The CG model is recovered at  $\theta = 1$. The EoS given in Eq. (\ref{gcg1}) with $\theta = 1$ was studied for the first time in 1904 by  Chaplygin in order to describe adiabatic processes \cite{gcg1}. The limiting case with $\theta \neq 1$ has been studied  in \cite{gcg3}. The idea that a cosmological model based on the Chaplygin gas could lead to the unification of DE and DM  was first proposed for $\theta = 1$ in the paper \cite{gcg4,gcg5}  and then generalized to $\theta \neq 1$ in \cite{Tawfik:2017ngn,gcg3}.

It was derived by Gorini  $\emph{et al.}$   \cite{gcg15} that the matter power spectrum is compatible with the observed one only if we have $\theta < 10^{-5}$, which means that the GCG is practically indistinguishable from the standard cosmological model with Cosmological Constant $\Lambda$. In \cite{gcg7}, the Chaplygin inflation has been studied in the framework of the  Loop Quantum Cosmology. Moreover, it was obtained that the parameters of the Chaplygin inflation model are consistent with the results of 5-year WMAP data \cite{Tawfik:2017ngn}.

The evolution of the energy density $\rho_D$ of the GCG model is given by
\begin{eqnarray}
    \rho_D = \left[D + \frac{B}{a^{3\left(\theta+1\right)}}\right]^{\frac{1}{\theta+1}},\label{gcg2}
\end{eqnarray}
where $B$ represents a constant of integration.

In principle, Eq. (\ref{gcg2}) admits a wide range of positive values of the parameter $\theta$,  however we must remember that it must ensure that the sound velocity (given by the relation $c^2_s=\frac{D\theta}{\rho^{\theta+1}}$) does not exceed the speed of the light $c$. Furthermore, as pointed out in Bento $\emph{et al.}$  \cite{gcg3},  only for the range of values  $0< \theta <1$  the analysis of the evolution of energy density fluctuations has a physical meaning.

We now want to reconstruct the potential and dynamics of this scalar field model in the framework of the DE model we are studying. The energy density $\rho_D$ and the pressure $p_D$ of the homogeneous and time-dependent scalar field
$\phi$ are given, respectively, by the following relations:
\begin{eqnarray}
\rho_D &=& \frac{1}{2}\dot{\phi}^2+V\left(  \phi \right), \label{gcg3}\\
p_D &=& \frac{1}{2}\dot{\phi}^2-V\left(  \phi \right). \label{gcg4}
\end{eqnarray}
Using the results of Eqs. (\ref{gcg3}) and (\ref{gcg4}), we derive the EoS parameter $\omega_D$ of the GCG model
\begin{eqnarray}
    \omega_D &=& \frac{\frac{1}{2} \dot{\phi}^2 - V\left(  \phi \right)}{ \frac{1}{2}\dot{\phi}^2 + V\left(\phi \right)}= \frac{ \dot{\phi}^2 - 2V\left(  \phi \right)}{ \dot{\phi}^2 + 2V\left(\phi \right)}. \label{gcg5}
\end{eqnarray}
Adding Eqs. (\ref{gcg3}) and (\ref{gcg4}), we can easily derive that the kinetic energy $\dot{\phi}^2$  term is given by the following relation:
\begin{eqnarray}
    \dot{\phi}^2 &=& \rho_D + p_D. \label{gcg6-1}
\end{eqnarray}
Using in Eq. (\ref{gcg6-1}) the definition of $p_D$ given in Eq. (\ref{gcg1}), we obtain the following relation:
\begin{eqnarray}
    \dot{\phi}^2 &=& \rho_D -\frac{D}{\rho_D^{\theta}}= \rho_D\left( 1-  \frac{D}{\rho_D^{\theta +1}}\right). \label{gcg6-11}
\end{eqnarray}
Considering the relation $\dot{\phi} = H \phi'$ and using the expression of $\rho_D$ given in Eq. (\ref{gcg2}), we can write the evolutionary form of $\phi$
\begin{eqnarray}
    \phi &=& \int _{a_0}^a\sqrt{3\Omega_D} \left(\sqrt{1-\frac{D}{ D+\frac{B}{a^{3\left(\theta+1\right)}}  }}\right)\frac{da}{a}, \label{gcg6-1newint}
\end{eqnarray}
whose solution, for a flat Dark Dominated Universe, is given by
\begin{eqnarray}
 \phi \left( a \right) &=& \frac{2\sqrt{3}}{3\left(1+\theta\right)}  \times \nonumber \\
 && \left[\log \left(a^{\frac{1}{2}(3+3 \theta )}\right)-\log \left(B+\sqrt{B\left(B+a^{3+3 \theta } D\right)}\right)\right]\label{murano4}.
\end{eqnarray}
Subtracting  Eqs. (\ref{gcg3}) and (\ref{gcg4}), we derive  the scalar potential $V\left(\phi \right)$ term
\begin{eqnarray}
 V\left(  \phi \right) &=& \frac{1}{2}\left( \rho_D - p_D\right) = \frac{\rho_D }{2}  \left( 1+ \frac{D}{\rho_D^{\theta +1}}\right), \label{gcg7-1}
\end{eqnarray}
where we used the definition of $p_D$ given in Eq. (\ref{gcg1}).

Considering the general expressions of $\rho_D$ given in Eq. (\ref{gcg2}), we obtain the following expression for  $V\left(  \phi \right)$
\begin{eqnarray}
V\left(  \phi \right) &=&  \frac{1}{2}\left[D + \frac{B}{a^{3\left(\theta+1\right)}}\right]^{\frac{1}{\theta+1}} +\frac{1}{2}\frac{D}{\left[D + \frac{B}{a^{3\left(\theta+1\right)}}\right]^{\frac{\theta}{\theta+1}}}. \label{gcg7-1new}
\end{eqnarray}
We now want to derive the general  expressions of the parameters $D$ and $B$ as functions of the other cosmological parameters.
Dividing Eq. (\ref{gcg1}) by $\rho_D$,  the EoS parameter  $\omega_D$ reads
\begin{eqnarray}
    \omega_D=   -\frac{D}{\rho_D^{\theta + 1}}, \label{gcg8}
\end{eqnarray}
which allows to derive the following expression for the parameter $D$
\begin{eqnarray}
    D= -\omega_D\rho_D^{\theta + 1}. \label{gcg9}
\end{eqnarray}
We also derive from Eq. (\ref{gcg2}) that the parameter $B$ can be written as
\begin{eqnarray}
B=a^{3\left(\theta + 1\right)} \left(\rho_D^{\theta + 1} - D   \right),\label{murano5}
\end{eqnarray}
which can be rewritten, substituting the expression of $D$ given in Eq. (\ref{gcg9}) as
\begin{eqnarray}
B=\left(a^3\rho_D\right)^{\theta+1}\left(1+\omega_D\right). \label{BBB}
\end{eqnarray}
Using in Eqs.  (\ref{gcg9}) and (\ref{BBB})  the expression of the EoS parameter $\omega_{Dpl}$ of the R-PLECHDE model given in Eq. (\ref{schirinzi}), we can write $D_{ pl}$ and $B_{ pl}$ as follows.
\begin{eqnarray}
D_{ pl}&=&\frac{ \rho_{D pl}^{\theta + 1}}{3} \left(\frac{1}{3c^2 -\delta R^{\gamma /2-1}}- \frac{1+\Omega_k}{\Omega_{D pl}} \right)  \nonumber \\
&=& \frac{ \rho_{D pl}^{\theta + 1}}{3} \left[\frac{1}{3c^2 -\delta R^{\gamma /2-1}}- \left(1+u_{pl}\right) \right], \label{murano6}\\
B_{ pl}&=&\left(a^3\rho_{D pl}\right)^{\theta+1}\times \nonumber\\
&&\left[1-\frac{1}{3}\left(\frac{1}{3c^2 -\delta R^{\gamma /2-1}}-\frac{1+\Omega_k}{\Omega_{D pl}}\right)\right]\nonumber \\
&=&   \left(a^3\rho_{D pl}\right)^{\theta+1}\times \nonumber\\
&&\left\{    1-\frac{1}{3}\left[\frac{1}{3c^2 -\delta R^{\gamma /2-1}}- \left(1+u_{pl}\right)\right]    \right\} . \label{murano7}
\end{eqnarray}

We now want to obtain the expressions of $D$ and $B$ for the limiting case of  a flat Dark Dominated Universe. Considering the expressions of $\rho_D$ and $\omega_D$ for the flat Dark Dominated case obtained, respectively, in Eqs. (\ref{rhoschi}) and (\ref{LEos}), we obtain that $D_{ Dark} $ and $B_{ Dark} $ are given by the following expressions:
\begin{eqnarray}
D_{ Dark} &=& \left[  3  \left(\frac{6c^2}{12c^2-1}\right)^2\left(\frac{1}{t^2}\right) \right]^{\theta + 1}\left( \frac{1-3c^2}{9c^2}  \right), \label{murano10}\\
B_{ Dark} &=& \left\{\left[ 3\left(\frac{6c^2}{12c^2-1}\right)^2 \right] t^{-2\left(3c^2 -1\right)/(12c^2-1)} \right\}^{\theta+1}\left(\frac{12c^2 -1}{9c^2}\right), \label{murano11}
\end{eqnarray}
where we used the expression of the scale factor given by the relation $a\left( t \right) = t^{6c^2/(12c^2-1)}$.

Using the value of $c^2$ found in the work of Gao, i.e., $c^2\approx 0.46$, we can write
\begin{eqnarray}
D_{ Dark} &\approx& -0.092\left(\frac{1.119}{t^2}\right)^{\theta + 1}, \label{murano10b}\\
B_{ Dark} &\approx &1.092 \left[\left( 1.119 \right) t^{-0.168} \right]^{\theta+1}. \label{murano11}
\end{eqnarray}

Furthermore, using the general definition of EoS parameter of DE $\omega_D$,  we can rewrite Eqs. (\ref{gcg6-1}) and (\ref{gcg7-1}) as
\begin{eqnarray}
    \dot{\phi}^2 &=& \left( 1+ \omega_D \right)\rho_D, \label{gcg6mura1}\\
    V\left(  \phi \right) &=& \frac{1}{2}\left( 1 - \omega_D \right)\rho_D. \label{gcg7mura2}
\end{eqnarray}
Using in Eqs. (\ref{gcg6mura1}) and (\ref{gcg7mura2}) the expression of the EoS parameter $\omega_{Dpl}$ of  the R-PLECHDE model given in Eq. (\ref{schirinzi}), we can derive the kinetic and the potential terms for the R-PLECHDE model
\begin{eqnarray}
    \dot{\phi}^2_{pl} &=& \rho_{Dpl}\left[ 1    -\frac{1}{3} \left(  \frac{1}{3c^2 -\delta R^{\gamma /2-1}}-  \frac{1+\Omega_k}{\Omega_{Dpl}}\right)    \right] \nonumber \\
 &&\rho_{Dpl}\left\{ 1    -\frac{1}{3} \left[  \frac{1}{3c^2 -\delta R^{\gamma /2-1}}-  \left(1+u_{pl}\right)\right]    \right\}, \label{gcg6mura1model1}\\
    V\left(  \phi \right)_{pl} &=& \frac{\rho_{Dpl}}{2}\left[ 1 +\frac{1}{3} \left(  \frac{1}{3c^2 -\delta R^{\gamma /2-1}}-  \frac{1+\Omega_k}{\Omega_{Dpl}}\right) \right]\nonumber \\
&& \frac{\rho_{Dpl}}{2}  \left\{  1   -\frac{1}{3}\left[ \frac{1}{3c^2 -\delta R^{\gamma /2-1}}- \left(1+u_{pl}\right)\right]   \right\}. \label{gcg7mura2model1}
\end{eqnarray}
We can obtain the evolutionary form of the GCG model integrating Eq. (\ref{gcg6mura1model1}) with respect to the scale factor $a\left(t\right)$
\begin{eqnarray}
	\phi\left(a\right)_{pl} - \phi\left(a_0\right)_{pl} &=& \int_{a_0}^{a}\sqrt{3\Omega_{Dpl}} \times \nonumber \\
&& \left[ 1    -\frac{1}{3} \left(  \frac{1}{3c^2 -\delta R^{\gamma /2-1}}-  \frac{1+\Omega_k}{\Omega_{Dpl}}\right)   \right]^{1/2} \frac{da}{a} \nonumber \\
&=& \int_{a_0}^{a}\sqrt{3\Omega_{Dpl}} \times \nonumber \\
&& \left\{  1   -\frac{1}{3}\left[ \frac{1}{3c^2 -\delta R^{\gamma /2-1}}- \left(1+u_{pl}\right)\right]   \right\}^{1/2} \frac{da}{a}, \label{gcg19}
\end{eqnarray}
where we used the relation $\dot{\phi}=\phi' H$.

In the limiting case for a flat Dark Dominated Universe, i.e., when $\Omega_{Dpl}=\Omega_{Dlog}=1$, $\Omega_k=\Omega_m=0$, and $\delta=0$ for the R-PLECHDE model, the scalar field and the potential of the GCG reduce, respectively, to
\begin{eqnarray}
\phi\left(t \right) &=&\left[ \frac{6c^2 }{\sqrt{3c^2  \left( 12c^2 -1\right)}}\right]\ln \left( t \right),\label{murano12} \\
V\left(t\right)&=& \frac{6c^2\left( 6c^2+1\right)}{\left(12c^2-1\right)^2}\left(\frac{1}{t^2}\right).\label{murano13gcg}
\end{eqnarray}
Using the value of $c^2$ found in the work of Gao, i.e., $c^2\approx 0.46$, we can write
\begin{eqnarray}
\phi\left(t \right) &\approx& 1.105\ln \left(t \right),\label{murano12} \\
V\left(t\right)&\approx& \left(\frac{0.508}{t^2}\right).\label{murano13gcg}
\end{eqnarray}
Moreover, after some algebraic calculations, we derive that the potential $V$ can be written as a function of the scalar field $\phi$ as
\begin{eqnarray}
V(\phi)&=&\left[\frac{6c^2(6c^2+1)}{(12c^2-1)^2}\right]e^{-\frac{\sqrt{3c^2(12c^2-1)}}{3c^2}\phi}.\label{murano14}
\end{eqnarray}
Using the value of $c^2$ found in the work of \cite{Gao} i.e., $c^2\approx 0.46$, we can write
\begin{eqnarray}
V(\phi)&\approx&0.508  e^{-1.810\phi}.\label{murano14}
\end{eqnarray}

\subsection{Modified  Chaplygin Gas (MCG) Model}

We now consider the second scalar field model considered in this paper, the Modified Chaplygin Gas (MCG) model.  This model represents a generalization of the GCG model  \cite{mcg1,mcg2,mcg3,mcg4} with the addition of a barotropic term and it is consistent with the 5-year WMAP data.

The MCG EoS is defined as follows \cite{mcg1,mcg5}.
\begin{equation}
p_D=A\rho_D-\frac{D}{\rho_D^\theta}, \label{mcg1}
\end{equation}
where $A$ and $D$ are two positive constant parameters and $\theta$ is constrained in the range $0 \leq \theta \leq 1$.
An interesting characteristic related to the MCG EoS is that it shows radiation era in the early phases of the Universe. At  late time, it behaves like a  model which can be fitted to a $\Lambda$CDM model.

The energy density $\rho_D$ of the MCG model is given by
\begin{equation}
\rho_D=\left[\frac{D}{A+1}+\frac{B}{a^{3(\theta+1)(A+1)}}\right]^{\frac{1}{\theta+1}},\label{mcg2}
\end{equation}
where $B$ represents a constant of integration.

We now want to reconstruct the potential and dynamics of the scalar field $\phi$. The energy density $\rho_D$ and pressure $p_D$ of the scalar field model are given, respectively, by
\begin{eqnarray}
\rho_D&=&\frac{1}{2}\dot\phi^2+V(\phi),\label{mcg3}\\
p_D&=&\frac{1}{2}\dot\phi^2-V(\phi).\label{mcg4}
\end{eqnarray}
Using the expressions of $\rho_D$ and $p_D$ given in Eqs. (\ref{mcg3}) and (\ref{mcg4}), we derive the EoS parameter $\omega_D$ for the MCG model
\begin{eqnarray}
\omega_D&=&  \frac{\frac{1}{2} \dot{\phi}^2-V(\phi)}{\frac{1}{2} \dot{\phi}^2+V(\phi)} = \frac{ \dot{\phi}^2-2V(\phi)}{ \dot{\phi}^2+2V(\phi)}.\label{mcg5}
\end{eqnarray}
Adding the expressions of Eqs. (\ref{mcg3}) and (\ref{mcg4}), we derive the kinetic energy $\dot{\phi}^2$ term
\begin{eqnarray}
    \dot{\phi}^2 &=& \rho_D + p_D. \label{mcg6-1}
\end{eqnarray}
Using the definition of $p_D$ given in Eq. (\ref{mcg4}), we obtain the following relation for $\dot{\phi}^2 $
\begin{eqnarray}
    \dot{\phi}^2 &=& \rho_D +A\rho_D-\frac{D}{\rho_D^\theta} = \rho_D\left( 1+ A - \frac{D}{\rho^{\theta +1}}    \right). \label{dotfi}
\end{eqnarray}
Inserting in Eq. (\ref{dotfi}) the expression of $\rho_D$ given in Eq. (\ref{mcg2}), we can write
\begin{eqnarray}
    \dot{\phi}^2 &=&  \rho_D\left\{ 1+A - \frac{D}{\left[\frac{D}{A+1}+\frac{B}{a^{3(\theta+1)(A+1)}}\right]}\right\}. \label{}
\end{eqnarray}
Using the relation $\dot{\phi} = H \phi'$, we can write
\begin{eqnarray}
\phi &=& \int_{a_0}^a  \sqrt{3\Omega_D} \sqrt{\left( 1+A\right) - \frac{D}{\left[\frac{D}{A+1}+\frac{B}{a^{3(\theta+1)(A+1)}}\right]} } \frac{da}{a}, \label{mcg6-1newint}
\end{eqnarray}
whose solution, for a flat Dark Dominated Universe, is given by
\begin{eqnarray}
\phi \left( a \right) &=&\frac{2}{\sqrt{3}\left( 1+A  \right)^{1/2}\left( 1+\theta \right)}\cdot \text{ArcTanh}\left[\frac{\sqrt{(1+A)
B+a^{3 (1+A) (1+\theta )} D}}{\sqrt{B\left(1+A\right)}}\right].
\end{eqnarray}
Subtracting  Eqs. (\ref{mcg3}) and (\ref{mcg4}), we can easily derive the scalar potential $V\left(\phi \right)$ term as
\begin{eqnarray}
 V\left(  \phi \right) &=& \frac{1}{2}\left( \rho_D - p_D\right). \label{mcg7-1}
\end{eqnarray}
Considering in Eq. (\ref{mcg7-1}) the general expressions of $p_D$ and $\rho_D$ given, respectively, in Eqs. (\ref{mcg1}) and (\ref{mcg2}), we can write   $V\left(\phi\right)$ as
\begin{eqnarray}
V\left(\phi\right) &=&\frac{\left(1-A\right)}{2} \left[\frac{D}{A+1}+\frac{B}{a^{3(\theta+1)(A+1)}}\right]^{\frac{1}{\theta+1}} +\frac{D}{2\left[\frac{D}{A+1}+\frac{B}{a^{3(\theta+1)(A+1)}}\right]^{\frac{\theta}{\theta+1}}}. \label{mcg7-1new}
\end{eqnarray}
We now want to find the expressions of the parameters $D$ and $B$. Dividing the expression of $\omega_D$ given in Eq. (\ref{mcg1}) by $\rho_D$,  we derive that the EoS parameter $\omega_D$ can be expressed as follows.
\begin{eqnarray}
\omega_D=A-\frac{D}{\rho_D^{\theta +1}}.\label{mcg8}
\end{eqnarray}
Therefore, from Eq. (\ref{mcg8}), we can easily obtain the following expression for $D$
\begin{eqnarray}
D=\rho_D^{\theta +1}\left(A-\omega_D\right). \label{mcg9}
\end{eqnarray}
Moreover, from Eq. (\ref{mcg8}), we can also derive the following expression for the parameter $A$:
\begin{eqnarray}
A = \omega_D+\frac{D}{\rho_D^{\theta +1}}.\label{mcg8new!}
\end{eqnarray}
Instead, from Eq. (\ref{mcg2}), we derive that $B$ can be rewritten as follows.
\begin{eqnarray}
B=a^{3(\theta +1)(A+1)}\left(\rho _{D}^{\theta+1}-\frac{D}{A+1}\right). \label{mcgb}
\end{eqnarray}
Substituting in Eq. (\ref{mcgb}) the expression of $D$ given in Eq (\ref{mcg9}), we obtain the following relation for $B$
\begin{eqnarray}
B=\left[a^{3\left(A+1\right)}\rho_D\right]^{1+\theta}  \left( \frac{1+\omega_D}{1+A} \right).\label{MMmcg}
\end{eqnarray}
Inserting in Eqs.  (\ref{mcg9}) and (\ref{MMmcg}) the expression of the EoS parameter $\omega_{Dpl}$ of the R-PLECHDE model given in Eq. (\ref{schirinzi}), we derive the following expressions for $D_{pl}$ and $B_{pl}$
\begin{eqnarray}
D_{ pl} &=&\left(\rho_{D pl}\right)^{\theta +1}\left [A +\frac{1}{3}\left(\frac{1}{3c^2 -\delta R^{\gamma /2-1}}-\frac{1+\Omega_k}{\Omega_{D pl}}\right) \right] \nonumber \\
&=&\left(\rho_{D pl}\right)^{\theta +1}\left\{     A +\frac{1}{3}\left[\frac{1}{3c^2 -\delta R^{\gamma /2-1}}-\left(1+u_{pl}\right)\right] \right\},\label{mcg11} \\
B_{ pl} &=& \frac{[a^{3(A+1)}\rho_{D pl}]^{\theta +1}}{1+A}\left[1-\frac{1}{3}\left( \frac{1}{3c^2 -\delta R^{\gamma /2-1}}-\frac{1+\Omega_k}{\Omega_{D pl}}  \right)\right] \nonumber \\
&=& \frac{[a^{3(A+1)}\rho_{D pl}]^{\theta +1}}{1+A}\left\{ 1-\frac{1}{3}\left[ \frac{1}{3c^2 -\delta R^{\gamma /2-1}}-\left(1+u_{pl}\right)  \right]\right\}.\label{mcg13}
\end{eqnarray}
In the limiting case of a flat Dark Dominated Universe, we obtain that the expressions of $D_{ Dark}$ and $B_{ Dark}$ are given, respectively, by
\begin{eqnarray}
D_{ Dark} &=&\left[3  \left(\frac{6c^2}{12c^2-1}\right)^2\left(\frac{1}{t^2}\right)\right]^{\theta+1}\left( A-\frac{1}{3}+\frac{1}{9c^2}  \right),\label{mcg11-3} \\
B_{ Dark} &=&\frac{\left[3  \left(\frac{6c^2}{12c^2-1}\right)^2t^{\frac{2\left(9A - 3c^2 +1   \right)}{12c^2 -1}}\right]^{\theta +1}}{1+A} \left( \frac{4}{3}-\frac{1}{9c^2}  \right),\label{mcg13-3}
\end{eqnarray}
where the results of Eqs. (\ref{rhoschi})  and (\ref{LEos}) for $\rho_D$ and $\omega_D$ are utilized.

Using the value  $c^2\approx 0.46$, we can write
\begin{eqnarray}
D_{ Dark} &\approx&\left[\left(\frac{1.119}{t^2}\right)\right]^{\theta+1}\left( A-0.092\right),\label{mcg11-3} \\
B_{ Dark} &\approx&\frac{1.092}{1+A}\left[1.119t^{0.442\left( 9A-0.38  \right)}\right]^{\theta +1} .\label{mcg13-3}
\end{eqnarray}
Furthermore, using the general definition of the EoS parameter $\omega_D$,  we can rewrite Eqs. (\ref{gcg6-1}) and (\ref{gcg7-1}) as follows.
\begin{eqnarray}
    \dot{\phi}^2 &=& \left( 1+ \omega_D \right)\rho_D, \label{gcg6mura1-1}\\
    V\left(  \phi \right) &=& \frac{1}{2}\left( 1 - \omega_D \right)\rho_D. \label{gcg7mura2-1}
\end{eqnarray}
Using in Eqs. (\ref{gcg6mura1-1}) and (\ref{gcg7mura2-1}) the expression of the EoS parameter $\omega_{Dpl}$ of  the R-PLECHDE model given in Eq. (\ref{schirinzi}), we can derive the kinetic and the potential terms for the R-PLECHDE model
\begin{eqnarray}
    \dot{\phi}^2_{pl} &=&  \rho_{Dpl}\left[ 1    -\frac{1}{3} \left(  \frac{1}{3c^2 -\delta R^{\gamma /2-1}}-  \frac{1+\Omega_k}{\Omega_{Dpl}}\right)    \right] \nonumber \\
 &&\rho_{Dpl}\left\{ 1  -\frac{1}{3} \left[ \frac{1}{3c^2 -\delta R^{\gamma /2-1}}-  \left(  1+u_{pl} \right)\right]  \right\}, \label{gcg6mura1model1-1}\\
    V\left(  \phi \right)_{pl} &=& \frac{\rho_{Dpl}}{2}\left[ 1 +\frac{1}{3} \left(  \frac{1}{3c^2 -\delta R^{\gamma /2-1}}-  \frac{1+\Omega_k}{\Omega_{Dpl}}\right) \right] \\
&&  \frac{\rho_{Dpl}}{2} \left\{ 1 +\frac{1}{3} \left[  \frac{1}{3c^2 -\delta R^{\gamma /2-1}}-  \left(  1+u_{pl} \right)\right]  \right\}  . \label{gcg7mura2model1-1}
\end{eqnarray}
We can obtain the evolutionary form of the GCG integrating Eq. (\ref{gcg6mura1model1-1}) with respect to the scale factor $a\left(t\right)$
\begin{eqnarray}
	\phi\left(a\right)_{pl} - \phi\left(a_0\right)_{pl} &=&	 \int_{a_0}^{a}\sqrt{3\Omega_{Dpl}} \times \nonumber \\
&& \left[ 1    -\frac{1}{3} \left(  \frac{1}{3c^2 -\delta R^{\gamma /2-1}}-  \frac{1+\Omega_k}{\Omega_{Dpl}}\right)   \right]^{1/2} \frac{da}{a} \nonumber \\
&=& \int_{a_0}^{a}\sqrt{3\Omega_{Dpl}} \times \nonumber \\
&& \left\{ 1   -\frac{1}{3}\left[ \frac{1}{3c^2 -\delta R^{\gamma /2-1}}- \left(1+u_{pl}\right)\right]   \right\}^{1/2} \frac{da}{a}, \label{gcg19-1}
\end{eqnarray}
where $\dot{\phi}=\phi' H$.

In the limiting case for a flat Dark Dominated Universe, i.e., when $\Omega_{Dpl}=1$, $\Omega_k=\Omega_m=0$ and $\delta=0$ for the R-PLECHDE model,
the scalar field and the potential of the GCG reduce, respectively, to
\begin{eqnarray}
\phi\left(t \right) &=& \frac{6c^2 }{\sqrt{3c^2  \left( 12c^2 -1\right)}}\ln \left( t \right),\label{murano12} \\
V\left(t\right)&=& \frac{6c^2\left( 6c^2+1\right)}{ \left(12c^2-1\right)^2}\left(\frac{1}{t^2}\right).\label{murano13gcg}
\end{eqnarray}
Using the value of $c^2$ found in the work of Gao, i.e., $c^2\approx 0.46$, we can write
\begin{eqnarray}
\phi\left(t \right) &\approx& 1.105\ln \left( t \right),\label{murano12} \\
V\left(t\right)&\approx& \left(\frac{0.508}{t^2}\right).\label{murano13gcg}
\end{eqnarray}
Moreover, after some algebraic calculations, we derive that the potential $V$ can be written as function of the scalar field $\phi$ as follows.
\begin{eqnarray}
V(\phi)&=&\left[\frac{6c^2(6c^2+1)}{(12c^2-1)^2}\right]e^{-\frac{\sqrt{3c^2(12c^2-1)}}{3c^2}\phi}.\label{murano14}
\end{eqnarray}
Using the value $c^2\approx 0.46$, we can write
\begin{eqnarray}
V(\phi)&\approx&0.508  e^{-1.810\phi}.\label{murano14}
\end{eqnarray}

\subsection{Modified Variable Chaplygin Gas (MVCG)}

We now consider the Modified Variable Chaplygin Gas (MVCG) model. Guo and Jhang \cite{mvcg1} recently proposed a model  known as Variable Chaplygin Gas (VCG)  with the following EoS
\begin{eqnarray}
p_D=- \frac{B}{\rho_D}, \label{murano16}
\end{eqnarray}
where $B$ indicates a function of the scale factor $a\left(t\right)$, i.e., $B=B\left( a\left(t\right)  \right)$. This particular assumption seems to be reasonable since it is related to scalar potential if CG is interpreted via Born-Infeld scalar field \cite{mvcg2}. In the following part of this Section, we will omit for simplicity the temporal dependence of the scale factor.  The VCG model  has been studied in the recent paper of \cite{mvcg3,mvcg4}. Debnath \cite{mvcg5} proposed the EoS of the Modified Variable Chaplygin Gas (MVCG) model in the following form.
\begin{eqnarray}
p_D=A\rho_D - \frac{B\left(a\right)}{\rho_D^{\theta}}. \label{murano17}
\end{eqnarray}
In this paper, we choose $B(a)= B_0 a^{-\delta_1}$. Therefore, we can write the pressure $p_D$ of the MVCG model as follows.
\begin{eqnarray}
p_D=A\rho_D - \frac{B_0a^{-\delta_1}}{\rho_D^{\theta}}. \label{murano18}
\end{eqnarray}
where $A$, $B_0$ and $\delta_1$ indicate three positive constant parameters, with $B_0$ being the present day value of $B$ and $\delta_1$ being the exponent of the scale factor. Moreover, $\theta$ is usually taken in the range of values $0\leq \theta \leq 1$.

In the limiting case of  $B_0= 0$, Eq. (\ref{murano18}) leads to a barotropic EoS (or equivalently to a barotropic fluid). In general, the barotropic EoS $p=A\rho$ is able to describe different kinds of media. For example, the limiting case with $A= -1$ (i.e., $p=-\rho$) leads to the Cosmological Constant case;  the limiting case with $A=-2/3$ leads to domain walls;  the limiting case with $A= -1/3$ produces cosmic strings;  the limiting case with $A= 0$ corresponds to dust or matter;  when  $A= 1/3$, we obtain the EoS for relativistic gas;  the limiting case with $A= 2/3$ gives the perfect gas;  finally, the limiting case with $A= 1$ represents the ultra-stiff matter. If we consider a constant expression of $B$, i.e., $B= B_0$, in Eq. (\ref{murano18}) (which is recovered for the limiting case of $\delta_1 = 0$), we recover the EoS of the original modified CG model. Eq. (\ref{murano18}) shows that, in the MVCG scenario, it interpolates between a radiation dominated phase $\left(A = \frac{1}{3}\right)$ and a quintessence-dominated phase described by a constant EoS. In the limiting case corresponding to $A=0$ and $\alpha =1$, we obtain the usual CG.  Recently, it was derived, using the latest Supernovae data, that models with $\alpha >1$ are also possible \cite{mvcg8}. We must also underline that, in the limiting case corresponding to $A=0$, Eq. (\ref{murano18}) yields a fluid with negative pressure which is generally characterized in the quintessence regime.

This modified form of the CG has also a phenomenological motivation since it can explain the flat rotational curves of galaxies \cite{mvcg9}. The galactic rotational velocity $V_c$ is related to the MVG parameter $A$ through the relation $V_c = \sqrt{2A}$ while the density parameter $\rho$ is related to the radial size of the galaxy $r$ by the relation $\rho = \frac{A}{2\pi G r^2}$. At high densities, the first term of the MVCG model is the dominant term and it produces the flat rotational curve that is consistent with present observations. The parameter $A$ varies from galaxy to galaxy due to the variations of $V_c$.

The energy density $\rho_D$ of the MVCG model is given by the following relation:
\begin{eqnarray}
\rho_D = \left\{\frac{3\left( \theta +1 \right)B_0}{\left[3\left( \theta +1 \right)\left(A+1\right)-\delta_1 \right]}\left(\frac{1}{a^{\delta_1}}\right)  - \frac{C}{a^{3\left( \theta +1 \right)\left(A+1\right)}} \right\}^{\frac{1}{1+\theta}},\label{murano19}
\end{eqnarray}
where $C$ is a positive constant of integration and $3\left( \theta +1 \right)\left(A+1\right)> \delta_1$ so that the first term results are positive. $\delta_1$ must be positive, otherwise the scale factor will tend to infinity, which implies that the energy density $\rho_D$ will tend to infinity too (which is not the case for the expanding Universe).

We now reconstruct the expressions of the potential and the dynamics of the scalar field. For this purpose, we consider a time dependent scalar field $\phi\left( t \right)$ with potential
$V\left(  \phi \right)$ , which are directly related with the energy density and pressure of MVCG as follows.
\begin{eqnarray}
\rho_D &=& \frac{1}{2}\dot{\phi}^2 + V\left(  \phi \right), \label{murano20} \\
p_D &=& \frac{1}{2}\dot{\phi}^2 - V\left(  \phi \right).\label{murano21}
\end{eqnarray}
Since the kinetic term is positive, we have that the MVCG is of quintessence type.

We know that the deceleration parameter $q$ can be expressed thanks to the following expression:
\begin{eqnarray}
q=-\frac{\ddot{a}}{aH^2}.\label{murano22}
\end{eqnarray}
In order to have an accelerating Universe, $q$ must be negative, i.e., we must have $\ddot{a}>0$ since the scale factor $a$ is positive and $H^2$ is always positive. $\ddot{a}>0$ implies the following relation:
\begin{eqnarray}
\left[\frac{2\left(1 + \theta\right) - \delta_1}{3\left(1 + \theta\right)\left(1 + A\right) - \delta_1}\right]a^{3\left( 1+\theta \right) \left( 1+A \right) -\delta_1} > \frac{C\left(1 + 3A\right)}{3B_0}.\label{murano23}
\end{eqnarray}
The result of Eq. (\ref{murano23})  requires $\delta_1 < 2\left(1 + \theta\right)$. Since we also have that $0 \leq \theta \leq 1$, we derive from the condition $\delta_1 < 2\left(1 + \theta\right)$ that the value of $\delta_1$ must be in the range $0 < \delta_1  < 4$.\\
This expression shows that for small value of scale factor we have decelerating Universe while for large values of the scale factor we have an accelerating Universe and the transition occurs when the scale factor assumes the value
\begin{eqnarray}
a= \left\{ \frac{C(1+3A)\left[3(1+\theta)(1+A)-\delta_1\right]}{3B_0\left[2(1+\theta)-\delta_1\right]}   \right\}^{\frac{1}{3(1+\theta)(1+A)-\delta_1}}.\label{murano24}
\end{eqnarray}
For small values of scale factor $a\left(t\right)$, we have the following relation between energy density $\rho$ and scale factor $a$
\begin{eqnarray}
\rho \cong \frac{C^{\frac{1}{1+\theta}}}{a^{3(1+A)}},\label{murano25}
\end{eqnarray}
which  corresponds to an Universe dominated by an EoS of the type $p=A\rho$.

Instead, for large values of the scale factor, we have the following relation between $\rho$ and the scale factor
\begin{eqnarray}
\rho \cong \left[ \frac{3(1 + \theta)B_0}{3(1 + \theta)(1 + A) - \delta_1}   \right]^{\frac{1}{(1+\theta)}}
a^{-\frac{\theta}{1+\theta}},\label{murano26}
\end{eqnarray}
which corresponds to the following EoS:
\begin{eqnarray}
p= \left[  \frac{\delta_1}{3(1 + \theta)} -1  \right] \rho,\label{murano27}
\end{eqnarray}
which describes a quintessence model \cite{mvcg6}.

We have that, in the limiting case corresponding to $\delta_1 = 0$, Eq. (\ref{murano27}) leads to the original modified Chaplygin gas scenario \cite{mvcg7}. However, Eq. (\ref{murano27}) shows that, in the variable modified Chaplygin gas scenario, it interpolates between a radiation dominated phase (which corresponds to the case with $A = 1/3$) and a quintessence-dominated phase described by the constant EoS $p =\gamma \rho$ where $\gamma = -1 + \frac{\delta_1}{3(1 + \theta)}< -\frac{1}{3}$.

We must also have that the energy density given in Eq. (\ref{murano19}) must be positive, so  that the scale factor $a\left( t \right)$ must obey the following condition:
\begin{eqnarray}
a\left( t \right) > \left\{ -\frac{C\left[3\left( \alpha +1 \right)\left(A+1\right)-\delta_1 \right]}{3\left( \alpha +1 \right)B_0}  \right\}^{\frac{1}{3\left( \alpha +1 \right)\left(A+1\right)-\delta_1}}. \label{murano28}
\end{eqnarray}
Therefore, we have that the minimum value of the scale factor $a\left( t\right)$ is given by
\begin{eqnarray}
a_{min}\left( t\right) = \left\{ -\frac{C\left[3\left( \theta +1 \right)\left(A+1\right)-\delta_1 \right]}{3\left( \theta+1 \right)B_0}  \right\}^{\frac{1}{3\left( \theta +1 \right)\left(A+1\right)-\delta_1}}. \label{murano29}
\end{eqnarray}
Adding  Eqs. (\ref{murano20}) and (\ref{murano21}), we can easily derive the kinetic energy $\dot{\phi}^2$  term as follows.
\begin{eqnarray}
    \dot{\phi}^2 &=& \rho_D + p_D. \label{murano30}
\end{eqnarray}
Instead, subtracting Eqs. (\ref{murano20}) and (\ref{murano21}), we can easily derive  the scalar potential $V\left(\phi \right)$ term as follows.
\begin{eqnarray}
 V\left(  \phi \right) &=& \frac{\left( \rho_D - p_D\right)}{2}. \label{murano31}
\end{eqnarray}
Inserting in Eq. (\ref{murano30}) the expression of $p_D$ and $\rho_D$ given, respectively, in Eqs. (\ref{murano18}) and (\ref{murano19}), we obtain the following expression for $\dot{\phi}^2$
\begin{eqnarray}
\dot{\phi}^2 &=& \left( 1+A\right) \left\{\frac{3\left( \theta +1 \right)B_0}{\left[3\left( \theta +1 \right)\left(A+1\right)-\delta_1 \right]}\left(\frac{1}{a^{\delta_1}}\right)  - \frac{C}{a^{3\left( \theta +1 \right)\left(A+1\right)}} \right\}^{\frac{1}{1+\theta}}\nonumber\\
 &-& \frac{B_0 a^{-\delta_1}}{\left\{ \frac{3\left( \theta +1 \right)B_0}{\left[3\left( \theta +1 \right)\left(A+1\right)-\delta_1 \right]}\left(\frac{1}{a^{\delta_1}}\right)  - \frac{C}{a^{3\left( \theta +1 \right)\left(A+1\right)}}  \right\}^{\frac{\theta}{1+\theta}}}.\label{murano32}
\end{eqnarray}
Using the relation   $\dot{\phi} = H \phi'$, we can write
\begin{eqnarray}
\phi &=& \int_{t_0}^t  \left( 1+A\right)^{1/2}  \left\{\frac{3\left( \theta +1 \right)B_0}{\left[3\left( \theta +1 \right)\left(A+1\right)-\delta_1 \right]}\left(\frac{1}{a^{\delta_1}}\right)  - \frac{C}{a^{3\left( \theta +1 \right)\left(A+1\right)}} \right\}^{\frac{1}{2\left(1+\theta\right)}}dt \nonumber \\
&-&\int_{t_0}^t  \frac{\left(B_0 a^{-\delta_1}\right)^{1/2}}{\left\{ \frac{3\left( \theta +1 \right)B_0}{\left[3\left( \theta +1 \right)\left(A+1\right)-\delta_1 \right]}\left(\frac{1}{a^{\delta_1}}\right)  - \frac{C}{a^{3\left( \theta +1 \right)\left(A+1\right)}}  \right\}^{\frac{\theta}{2\left(1+\theta\right)}}}dt.   \label{murano33}
\end{eqnarray}
From the Friedmann equation given in Eq. (\ref{7}), for $k = 0$ (i.e., for a flat Universe), we get the explicit form of $t$ as function of the scale factor $a\left( t  \right)$ as follows.
\begin{eqnarray}
t=Ka^{\frac{\delta_1}{2\left(1+\theta   \right)}}\,_2F_1\left[ \frac{1}{2\left(1+\theta \right)}, -z,1-z,-\left(\frac{C}{K}\right)a^{-\frac{\delta_1}{2\left(1+\theta   \right)z}} \right],\label{murano34}
\end{eqnarray}
where the quantities $K$ and $z$ are defined as follows.
\begin{eqnarray}
K&=&\frac{2}{\delta_1}\left[\left( 1+\theta \right)^{\theta} \sqrt{\frac{\delta_1}{6B_0z}}   \right]^{\frac{1}{1+\theta}}, \label{murano35}\\
z &=& \frac{\delta_1}{2\left( 1+\theta \right)\left[3\left( 1+A  \right)\left( 1+\theta \right)-\delta_1   \right]}.\label{murano36}
\end{eqnarray}
Moreover, $_2F_1$ represents the hypergeomtric function of second type.

For a flat Universe, considering the expressions of $t$ given in Eq. (\ref{murano34}), we derive the following expression of $\phi$:
\begin{eqnarray}
\phi &=& \frac{\sqrt{1+A}}{3\left( 1+A  \right)\left( 1+\theta \right)-\delta_1  }\times \nonumber \\
&&\left\{2\log \left( \sqrt{u+x} + \sqrt{u+y}  \right) - \sqrt{\frac{y}{x}}\log \left[ \frac{\left( \sqrt{x\left(u+x\right)} + \sqrt{y\left(u+y\right)}  \right)^2}{x^{3/2}\sqrt{y}u}  \right]     \right\},\label{murano37}
\end{eqnarray}
where the parameters $x$, $y$ and $u$ are defined, respectively, as follows.
\begin{eqnarray}
x &=& \frac{\delta_1}{1+A}, \label{murano38} \\
y &=&3\left(1+\theta \right), \label{murano39}\\
u &=& \left(\frac{\delta_1 C}{B_0}\right)a^{\delta_1\left( 1-\frac{y}{x}  \right)}.\label{murano40}
\end{eqnarray}
Moreover, inserting in Eq. (\ref{murano31}) the expression of $p_D$ and $\rho_D$ given, respectively, in Eqs. (\ref{murano18}) and (\ref{murano19}), we obtain the following expression for $V\left(\phi\right)$:
\begin{eqnarray}
V\left(\phi\right) &=& \frac{\left( 1-A\right)}{2}\left\{ \frac{3\left( \theta +1 \right)B_0}{\left[3\left( \theta +1 \right)\left(A+1\right)-\delta_1 \right]}\left(\frac{1}{a^{\delta_1}}\right)  - \frac{C}{a^{3\left( \theta +1 \right)\left(A+1\right)}}  \right\}^{\frac{1}{1+\theta}} \nonumber\\
 && +\frac{B_0 a^{-\delta_1}}{2\left\{  \frac{3\left( \theta +1 \right)B_0}{\left[3\left( \theta +1 \right)\left(A+1\right)-\delta_1 \right]}\left(\frac{1}{a^{\delta_1}}\right)  - \frac{C}{a^{3\left( \theta +1 \right)\left(A+1\right)}}  \right\}^{\frac{\theta}{1+\theta}}}.\label{murano40}
\end{eqnarray}
We now want to derive the general expressions of the parameters $B_0$ and $C$.

Dividing the expression of $p_D$ given in Eq. (\ref{murano18}) by $\rho_D$, we obtain the following relation for the EoS parameter $\omega_D$
\begin{eqnarray}
\omega_D=A - \frac{B_0a^{-\delta_1}}{\rho_D^{\theta+1}},\label{murano41}
\end{eqnarray}
which leads to the following expression of $B_0$
\begin{eqnarray}
B_0 =a^{\delta_1}\left( A- \omega_D  \right)\rho_D ^{\theta +1}.\label{murano42}
\end{eqnarray}
Instead, from the expression of $\rho_D$ given in Eq. (\ref{murano19}), we find that
\begin{eqnarray}
C =\left\{\frac{3\left( \theta +1 \right)B_0}{\left[3\left( \theta +1 \right)\left(A+1\right)-\delta_1 \right]}\left(\frac{1}{a^{\delta_1}}\right) - \rho_D^{1+\theta} \right\} a^{-3\left( \theta +1 \right)\left(A+1\right)}.\label{murano43}
\end{eqnarray}
Using in Eq. (\ref{murano43}) the expression of $B_0$ obtained in Eq. (\ref{murano42}), we obtain the following expression for $C$
\begin{eqnarray}
C= \left[\rho_D  a^{-3\left(A+1\right)}\right]^{\theta +1}\left[\frac{3\left(\theta+1 \right)\left(A - \omega_D\right)}{3\left(\theta+1 \right)\left( A+1  \right)-\delta_1}-1     \right].\label{murano44}
\end{eqnarray}
Substituting  in Eqs. (\ref{murano42}) and (\ref{murano44}) the expression of the EoS parameter for the R-PLECHDE model given in Eq. (\ref{schirinzi}), we find that $B_{0,pl}$ and $C_{pl}$ can be written as follows.
\begin{eqnarray}
B_{0, pl} &=&a^{\delta_1}\left[ A +\frac{1}{3}\left(\frac{1}{3c^2 -\delta R^{\gamma /2-1}}-\frac{1+\Omega_k}{\Omega_{D pl}}\right)   \right]  \rho_{D pl} ^{\theta +1} \nonumber \\
&=&a^{\delta_1}\left\{ A +\frac{1}{3}\left[ \frac{1}{3c^2 -\delta R^{\gamma /2-1}}-\left(1+u_{pl}\right)\right]   \right\}    \rho_{D pl} ^{\theta +1},\label{murano45}\\
C_{ pl}&=& \left[\rho_{D pl}  a^{-3\left(A+1\right)}\right]^{\theta +1}\left\{\frac{3\left(\theta+1 \right)\left[A  +\frac{1}{3}\left(\frac{1}{3c^2 -\delta R^{\gamma /2-1}}-\frac{1+\Omega_k}{\Omega_{D pl}}\right)  \right]}{3\left(\theta+1 \right)\left( A+1  \right)-\delta_1}-1     \right\} \nonumber \\
&=& \left[\rho_{D pl}  a^{-3\left(A+1\right)}\right]^{\theta +1}\times \nonumber \\
&&\left\{\frac{3\left(\theta+1 \right)\left[A  +\frac{1}{3}\left(\frac{1}{3c^2 -\delta R^{\gamma /2-1}}-\left(1+u_{pl}\right)\right)  \right]}{3\left(\theta+1 \right)\left( A+1  \right)-\delta_1}-1 \right\}.\label{murano46}
\end{eqnarray}

In the limiting case of a flat Dark Dominated Universe, we have that $B_{0,Dark}$ and $C_{Dark}$ can be written as follows.
\begin{eqnarray}
B_{0, Dark} &=&\left[3  \left(\frac{6c^2}{12c^2-1}\right)^2\left(\frac{1}{t^2}\right)\right]^{\theta +1}t^{6\delta_1 c^2/(12c^2-1)}\left( A - \frac{1}{3}+ \frac{1}{9c^2}   \right),\label{murano49}\\
C_{ Dark}&=&\left\{\left[  3 \left(\frac{6c^2}{12c^2-1}\right)^2 \right]t^{-18\left(A+1\right)c^2/(12c^2-1)+2}\right\}^{\theta +1}\times \nonumber \\
&&\left[\frac{3\left(\theta +1 \right)\left(A - \frac{1}{3}+ \frac{1}{9c^2} \right)}{3\left(\theta +1 \right)\left( A+1  \right)-\delta_1}-1     \right],
 \label{murano50}
\end{eqnarray}
where the results of Eqs. (\ref{rhoschi}) and (\ref{LEos}) for $\rho_D$ and $\omega_D$ are utilized.

Using the value $c^2\approx 0.46$, we can write
\begin{eqnarray}
B_{0, Dark} &\approx&\left[\left(\frac{1.118568}{t^2}\right)\right]^{\theta +1}t^{1.105 \delta_1 }\left( A - 0.092  \right),\label{murano49}\\
C_{ Dark}&\approx&\left[\left( 1.118568 \right)t^{-1.831867A + 0.16184}\right]^{\theta +1}\times \nonumber \\
&&\left[\frac{3\left(\theta +1 \right)\left(A -0.092  \right)}{3\left(\theta +1 \right)\left( A+1  \right)-\delta_1}-1 \right].
 \label{murano50}
\end{eqnarray}

Furthermore, using the general definition of EoS parameter, we can rewrite Eqs. (\ref{gcg6-1}) and (\ref{gcg7-1}) as follows.
\begin{eqnarray}
    \dot{\phi}^2 &=& \left( 1+ \omega_D \right)\rho_D, \label{gcg6mura1-1-1}\\
    V\left(  \phi \right) &=& \frac{1}{2}\left( 1 - \omega_D \right)\rho_D. \label{gcg7mura2-1-1}
\end{eqnarray}
Using in Eqs. (\ref{gcg6mura1-1-1}) and (\ref{gcg7mura2-1-1}) the expression of the EoS parameter $\omega_{Dpl}$ of  the R-PLECHDE model given in Eq. (\ref{schirinzi}), we can derive the kinetic and the potential terms for the R-PLECHDE model
\begin{eqnarray}
    \dot{\phi}^2_{pl}  &=& \rho_{Dpl}\left[ 1    -\frac{1}{3} \left(  \frac{1}{3c^2 -\delta R^{\gamma /2-1}}-  \frac{1+\Omega_k}{\Omega_{Dpl}}\right)    \right] \nonumber \\
&& \rho_{Dpl}\left\{ 1    -\frac{1}{3} \left[  \frac{1}{3c^2 -\delta R^{\gamma /2-1}}-  \left( 1+u_{pl}  \right)   \right]    \right\}, \label{gcg6mura1model1-1-1}\\
    V\left(  \phi \right)_{pl}  &=& \frac{\rho_{Dpl}}{2}\left[ 1 +\frac{1}{3} \left(  \frac{1}{3c^2 -\delta R^{\gamma /2-1}}-  \frac{1+\Omega_k}{\Omega_{Dpl}}\right) \right] \nonumber \\
&& \frac{\rho_{Dpl}}{2}\left\{  1 +\frac{1}{3} \left[  \frac{1}{3c^2 -\delta R^{\gamma /2-1}}-  \left(  1+u_{pl} \right)  \right] \right\}. \label{gcg7mura2model1-1-1}
\end{eqnarray}
We can obtain the evolutionary form of the GCG model integrating Eq. (\ref{gcg6mura1model1-1-1}) with respect to the scale factor $a\left(t\right)$
\begin{eqnarray}
	\phi\left(a\right)_{pl} - \phi\left(a_0\right)_{pl}  &=&	 \int_{a_0}^{a}\sqrt{3\Omega_{Dpl}} \times \nonumber \\
&& \left[ 1    -\frac{1}{3} \left(  \frac{1}{3c^2 -\delta R^{\gamma /2-1}}-  \frac{1+\Omega_k}{\Omega_{Dpl}}\right)   \right]^{1/2} \frac{da}{a} \nonumber \\
&=& \int_{a_0}^{a}\sqrt{3\Omega_{Dpl}} \times \nonumber \\
&& \left\{ 1   -\frac{1}{3}\left[ \frac{1}{3c^2 -\delta R^{\gamma /2-1}}- \left(1+u_{pl}\right)\right]   \right\}^{1/2} \frac{da}{a}, \label{gcg19-1-1}
\end{eqnarray}
where $\dot{\phi}=\phi' H$.

In the limiting case for a flat Dark Dominated Universe, i.e., when $\Omega_{Dpl}=\Omega_{Dlog}=1$, $\Omega_k=\Omega_m=0$ and $\delta=0$ for the R-PLECHDE model, the scalar field $\phi$ and the potential $V$ of the GCG model reduce, respectively, to
\begin{eqnarray}
\phi\left(t \right) &=& \left[\frac{6c^2 }{\sqrt{3c^2  \left( 12c^2 -1\right)}}\right]\ln \left( t \right),\label{murano12} \\
V\left(t\right)&=& \frac{6c^2\left( 6c^2+1\right)}{ \left(12c^2-1\right)^2}\left(\frac{1}{t^2}\right).\label{murano13gcg}
\end{eqnarray}
Using the value $c^2\approx 0.46$, we can write
\begin{eqnarray}
\phi\left(t \right) &\approx& 1.105\ln \left( t \right),\label{murano12} \\
V\left(t\right)&\approx& \left(\frac{0.508}{t^2}\right).\label{murano13gcg}
\end{eqnarray}
Moreover, after some algebraic calculations, we derive that the potential $V$ as function of the scalar field $\phi$
\begin{eqnarray}
V(\phi)&=&\left[\frac{6c^2(6c^2+1)}{(12c^2-1)^2}\right]e^{-\frac{\sqrt{3c^2(12c^2-1)}}{3c^2}\phi}.\label{murano14}
\end{eqnarray}
Using the value $c^2\approx 0.46$, we obtain
\begin{eqnarray}
V(\phi)&\approx&0.508  e^{-1.810\phi}. \label{murano14}
\end{eqnarray}

\subsection{New Modified Chaplygin Gas (NMCG) Model}

We now consider as model that represents the DE the New Modified Chaplygin Gas (NMCG), which has EoS given by \cite{newm1}
\begin{eqnarray}
p_D = B \rho_D - \frac{K\left( a\right)}{\rho_D^{\theta}},\label{murano55}
\end{eqnarray}
where $K\left( a\right)$ is a function of the scale factor $a$, $B$ is a positive constant and $\theta$ assumes values in the range $0\leq \theta \leq 1$. If we take $K\left(a\right)$ in the form  $K\left(a\right) = -\omega_DA_1a^{-3\left(\omega_D+1\right)\left(\theta +1 \right)}$  as introduced by  \cite{newm2}, Eq. (\ref{murano55}) can be written as follows.
\begin{eqnarray}
p_D = B \rho_D +\left( \frac{\omega_DA_1}{\rho_D^{\theta}}\right)a^{-3\left(\omega_D+1\right)\left(\theta +1 \right)}.\label{murano56}
\end{eqnarray}
The energy density $\rho_D$ of the NMCG model  is given by
\begin{eqnarray}
\rho_D
= \left[\left(\frac{\omega_DA_1}{\omega_D-B}\right)a^{-3\left(\omega_D+1\right)\left(\theta +1 \right)}+ B_1a^{-3\left(B+1\right)\left(\theta +1 \right)} \right]^{\frac{1}{1+\theta}},\label{murano57}
\end{eqnarray}
where $B_1$ is a constant of integration.

We now want to derive the expressions of the parameters $B_1$ and $A_1$.
From Eq. (\ref{murano57}), we can easily derive the following expression for $B_1$:
\begin{eqnarray}
B_1 = a^{3\left(B+1\right)\left(\theta+1\right)}\left[ \rho_D^{\theta+1}-
\left( \frac{\omega_D}{\omega_D-B} \right)   A_1a^{-3\left(\omega_D+1\right)\left(\theta+1\right)} \right]. \label{murano58}
\end{eqnarray}
Moreover,  dividing by $\rho_D$ the expression of $p_D$ given in Eq. (\ref{murano56}) and using the definition of EoS parameter $\omega_D$, we obtain
\begin{eqnarray}
A_1 = \left(\frac{\omega_D-B}{\omega_D}\right)\rho_D^{\theta +1}a^{3\left(\omega_D+1\right)\left(\theta +1\right)}.\label{murano59}
\end{eqnarray}
We can obtain the final expressions of $A_1$ and $B_1$ for the R-PLECHDE model inserting in Eqs. (\ref{murano58}) and (\ref{murano59}) the expression of the EoS parameter given in Eq. (\ref{schirinzi})
\begin{eqnarray}
B_{1, pl} &=& a^{3\left(B+1\right)\left(\theta+1\right)}\left[ \rho_{Dpl}^{\theta+1}-
\left( \frac{\omega_{Dpl}}{\omega_{Dpl}-B}\right)    A_1a^{-3\left(\omega_{Dpl}+1\right)\left(\theta+1\right)} \right], \label{murano58pl}\\
 A_{1, pl} &=&\left( \frac{\omega_{Dpl}-B}{\omega_{Dpl}}\right)\rho_{Dpl}^{\theta +1}a^{3\left(\omega_{Dpl}+1\right)\left(\theta +1\right)}.\label{murano59pl}
\end{eqnarray}
In the limiting case of a flat Dark Dominated Universe, we obtain
\begin{eqnarray}
B_{1, Dark} &=& t^{2\left( \theta + 1\right)}\left\{ \left[ 3  \left(\frac{6c^2}{12c^2-1}\right)^2  \right]^{\theta+1} - \left[\frac{3c^2 -1}{3c^2\left( 1-3B \right)-1}\right]    A_1\right\}, \label{murano65}\\
A_{1, Dark} &=& \frac{3c^2\left( 1-3B \right)-1}{3c^2 -1} \left[ 3  \left(\frac{6c^2}{12c^2-1}\right)^2 \right]^{\theta +1},\label{murano66}
\end{eqnarray}
where the results of Eqs. (\ref{rhoschi}) and (\ref{LEos}) for $\rho_D$ and $\omega_D$ are utilized.

For  $c^2\approx 0.46$, we obtain
\begin{eqnarray}
B_{1, Dark} &\approx& t^{2\left( \theta + 1\right)}\left[ \left( 1.119 \right)^{\theta+1} + \frac{0.38A_1}{4.14B - 0.38}    \right], \label{murano65}\\
A_{1, Dark} &\approx& -\left(\frac{4.14B - 0.38}{0.38}\right)  \left( 1.119\right)^{\theta +1}.\label{murano66}
\end{eqnarray}
Furthermore, using the general definition of EoS parameter,  we can rewrite Eqs. (\ref{gcg6-1}) and (\ref{gcg7-1}) as follows.
\begin{eqnarray}
    \dot{\phi}^2 &=& \left( 1+ \omega_D \right)\rho_D, \label{gcg6mura1-1-1-1}\\
    V\left(  \phi \right) &=& \frac{1}{2}\left( 1 - \omega_D \right)\rho_D. \label{gcg7mura2-1-1-1}
\end{eqnarray}
Using in Eqs. (\ref{gcg6mura1-1-1}) and (\ref{gcg7mura2-1-1-1}) the expression of the EoS parameter $\omega_{Dpl}$ of  the R-PLECHDE model given in Eq. (\ref{schirinzi}), we can derive the kinetic and the potential terms for the R-PLECHDE model
\begin{eqnarray}
    \dot{\phi}^2_{pl} &=& \rho_{Dpl}\left[ 1    -\frac{1}{3} \left(  \frac{1}{3c^2 -\delta R^{\gamma /2-1}}-  \frac{1+\Omega_k}{\Omega_{Dpl}}\right)    \right] \nonumber \\
&& \rho_{Dpl}\left\{ 1    -\frac{1}{3} \left[  \frac{1}{3c^2 -\delta R^{\gamma /2-1}}-  \left(  1+u_{pl}  \right)  \right]    \right\}, \label{upupu}\\
    V\left(  \phi \right)_{pl}  &=& \frac{\rho_{Dpl}}{2}\left[ 1 +\frac{1}{3} \left(  \frac{1}{3c^2 -\delta R^{\gamma /2-1}}-  \frac{1+\Omega_k}{\Omega_{Dpl}}\right) \right] \nonumber \\
&& \frac{\rho_{Dpl}}{2}\left\{ 1 +\frac{1}{3} \left[  \frac{1}{3c^2 -\delta R^{\gamma /2-1}}-    \left( 1+u_{pl}  \right) \right] \right\} . \label{gcg7mura2model1-1-1-1}
\end{eqnarray}
We can obtain the evolutionary form of the GCG integrating Eq. (\ref{upupu}) with respect to the scale factor $a\left(t\right)$
\begin{eqnarray}
	\phi\left(a\right)_{pl}  - \phi\left(a_0\right)_{pl}  &=&	 \int_{a_0}^{a}\sqrt{3\Omega_{Dpl}} \times \nonumber \\
&& \left[ 1    -\frac{1}{3} \left(  \frac{1}{3c^2 -\delta R^{\gamma /2-1}}-  \frac{1+\Omega_k}{\Omega_{Dpl} }\right)   \right]^{1/2} \frac{da}{a} \nonumber \\
&=& \int_{a_0}^{a}\sqrt{3\Omega_{Dpl}} \times \nonumber \\
&& \left\{ 1   -\frac{1}{3}\left[ \frac{1}{3c^2 -\delta R^{\gamma /2-1}}- \left(1+u_{pl}\right)\right]   \right\}^{1/2} \frac{da}{a}, \label{gcg19-1-1-1}
\end{eqnarray}
where $\dot{\phi}=\phi' H$.

In the limiting case for a flat Dark Dominated Universe, i.e., when $\Omega_{Dpl}=\Omega_{Dlog}=1$, $\Omega_k=\Omega_m=0$, $\delta=0$ for the R-PLECHDE model, the scalar field and the potential of the GCG reduce, respectively, to
\begin{eqnarray}
\phi\left(t \right) &=& \frac{6c^2 }{\sqrt{3c^2  \left( 12c^2 -1\right)}}\ln \left( t \right),\label{murano12} \\
V\left(t\right)&=& \frac{6c^2\left( 6c^2+1\right)}{ \left(12c^2-1\right)^2}\left(\frac{1}{t^2}\right).\label{murano13gcg}
\end{eqnarray}
For $c^2\approx 0.46$, we obtain
\begin{eqnarray}
\phi\left(t \right) &\approx& 1.105\ln \left( t \right),\label{murano12} \\
V\left(t\right)&\approx& \left(\frac{0.508}{t^2}\right).\label{murano13gcg}
\end{eqnarray}
Moreover, after some algebraic calculations, we derive that the potential $V$ can be written as function of the scalar field $\phi$ as follows.
\begin{eqnarray}
V(\phi)&=&\left[\frac{6c^2(6c^2+1)}{(12c^2-1)^2}\right]e^{-\frac{\sqrt{3c^2(12c^2-1)}}{3c^2}\phi}.\label{murano14}
\end{eqnarray}
For $c^2\approx 0.46$, we have
\begin{eqnarray}
V(\phi)&\approx&0.508  e^{-1.810\phi}.\label{murano14}
\end{eqnarray}

\subsection{Viscous Generalized Chaplygin Gas (VGCG) Model}

We now consider the Viscous Generalized Chaplygin Gas (VGCG) Model. In order to be more general possible respect to the previous sections, we now consider a viscous DE interacting with DM. In an isotropic and homogeneous FRW Universe, the dissipative effects are generated by  the presence of bulk viscosity in the cosmic fluids \cite{Tawfik:2021rvv,Tawfik:2011gh,Tawfik:2011sh,Tawfik:2011mw,Tawfik:2010bm,Tawfik:2009mk}. Some dissipative processes in the Universe (including bulk viscosity, shear viscosity and heat transport) are believed to be present in any realistic theory of the evolution of the Universe and they have been widely studied in  \cite{visc29,visc30,visc31,visc32}. The role of viscosity has been widely discussed and it is considered a promising candidate which can explain several cosmological problems like DE. The viscous DE model can provide an explanation the high photon to baryon ratio \cite{visc18} and it leads to an inflationary scenario in the early phase of the Universe evolution \cite{visc19}. The coefficient of viscosity should decrease as the Universe expands;  moreover, its presence can explain the current accelerated expansion of the Universe \cite{visc20,visc21,visc22}. This model is also consistent with astrophysical observations at  lower redshifts, and a viscous cosmic model favors a standard cold DM model with Cosmological Constant ($\Lambda$CDM) in the later cosmic evolution \cite{visc25}. The model also presents the scenario of phantom crossing \cite{visc26}.

The general theory of dissipation in a relativistic imperfect fluid was done  by Eckart \cite{visc33} and, in a different formulation, by Landau and Lifshitz \cite{visc34}. This is only the first order deviation from equilibrium and may has a causality problem \cite{Tawfik:2021rvv,Tawfik:2011gh,Tawfik:2011sh,Tawfik:2011mw,Tawfik:2010bm,Tawfik:2009mk}. The full causal theory was developed by Israel $\&$ Stewart \cite{visc36} and it has also been studied in the evolution of the early Universe \cite{visc37}. However, the character of the evolution equation is very complicated in the full causal theory \cite{Tawfik:2021rvv,Tawfik:2011gh,Tawfik:2011sh,Tawfik:2011mw,Tawfik:2010bm,Tawfik:2009mk}. Fortunately, since the phenomena are quasi-stationary, i.e., they slowly vary on space and time scale which are characterized by the mean free path and the mean collision time of the fluid particles, the conventional theory can be still considered valid. In the case of an isotropic and homogeneous FRW Universe, the dissipative process can be modeled as a bulk viscosity within a thermodynamical approach, while the shear viscosity  can be safely ignored \cite{visc40}. For other  works on viscous DE models see \cite{visc41,visc43,visc44,visc45,visc46,visc48,visc49,visc50}.

DE with bulk viscosity has the peculiar property to cause an accelerated expansion of phantom-type in late evolution stages of the Universe \cite{brevik1,brevik2,brevik3} and it can also alleviate several cosmological problems like the coincidence and the age problems and also the phantom crossing.

The observations also indicate that the Universe medium is not a perfect fluid and the viscosity is concerned in the evolution of the Universe (for more details see \cite{ren1} and references therein). In the framework of FRW metric, the shear viscosity has no contribution in the energy-momentum tensor $T^{\mu \nu}$ and the  bulk viscosity behaves like an effective pressure \cite{ren2}.

The bulk viscosity introduces dissipation by only redefining the effective pressure $p_{eff}$ given by the relation $p_{eff} = p-3\nu H$,  where $\nu$ indicates the bulk viscosity coefficient. The condition $\nu > 0$ guaranties a positive entropy production, consequently, no violation of the second law of the thermodynamics \cite{visc51}. The case $\nu = \tau H$, implying the bulk viscosity is proportional to the fluid velocity vector, is physically natural and has been considered earlier in a astrophysical context, see  \cite{visc52}.

The energy conservation equation yields the following expression of the energy density of the VGCG model $\rho_D$ \cite{visc9}
\begin{eqnarray}
\rho_D = \left[  \frac{Da^{-3\left(\theta+1 \right)\left(1- \nu \varrho \right)}-\chi}{1- \nu \varrho }  \right]^{\frac{1}{\theta +1}},\label{murano72}
\end{eqnarray}
where $\varrho  = m_p^{-1}\sqrt{1-r_m}$ (with $r_m = \frac{\rho_m}{\rho_D} = \frac{\Omega_m}{\Omega_D}$) and $D$ is constant of integration.

The energy-momentum tensor corresponding to the bulk viscous fluid is given by
\begin{eqnarray}
T = \left(  \rho+ \bar{p}  \right)u_{\mu}u_{\nu} - \bar{p}g_{\mu \nu},\label{murano73}
\end{eqnarray}
where
\begin{eqnarray}
\bar{p} = p_D -3\varepsilon H \label{murano74}
\end{eqnarray}
represents the total pressure which involves the proper pressure $p$, the bulk viscosity coefficient $\varepsilon$  and the Hubble parameter $H$.

We have that, in this case, $p_D = \frac{\chi}{\rho_D^{\theta}}$, with $\chi>0$. We notice that the first term on the right hand side of Eq. (\ref{murano74}) mimics the GCG model and the parameter $\theta$ varies in the range $0<\theta <1$. If $\theta =1$, it leads the Chaplygin gas model otherwise,  if  $\theta<0$, it corresponds to a polytropic gas.

We choose an expression of $\varepsilon$ which depends on the energy density $\rho_D$ in the following way: $\varepsilon = \nu \rho_D^{1/2} $ (with $\nu$ being a constant parameter).
Therefore, using in Eq. (\ref{murano74}) the expression of $\varepsilon$ we have chosen, we can rewrite $\bar{p}$ as follows.
\begin{eqnarray}
\bar{p} = \frac{\chi}{\rho_D^{\theta}} -3\nu H\rho_D^{1/2}.  \label{murano74new}
\end{eqnarray}
The energy density $\rho_D$ and pressure $p_D$ of the viscous dark energy model are given by the following expressions:
\begin{eqnarray}
\rho_D &=& \left[  \frac{Da^{-3\left(\theta+1 \right)\left(1- \nu \varrho \right)}-\chi}{1- \nu \varrho }  \right]^{\frac{1}{\theta +1}}, \label{schiri1}\\
p_D &=& \chi \left[  \frac{1- \nu \varrho } {Da^{-3\left(\theta+1 \right)\left(1- \nu \varrho \right)}-\chi} \right]^{\frac{\theta}{\theta +1}}-3\nu H \left[  \frac{Da^{-3\left(\theta+1 \right)\left(1- \nu \varrho \right)}-\chi}{1- \nu \varrho }  \right]^{1/2}.\label{schiri2}
\end{eqnarray}
We now want to derive the expressions of the parameters $\chi$ and $D$. The expression of $\chi$ as function of the EoS parameter $\omega_D$ can be easily derived from Eq. (\ref{murano74new}). Dividing  the result of Eq. (\ref{murano74new}) by $\rho_D$ and using the general definition of $\omega_D$, we obtain
\begin{eqnarray}
\chi = \rho_D^{\theta +1}\left( 3\nu H \rho_D^{-1/2}+ \omega_D  \right).\label{murano75}
\end{eqnarray}
We can determine the expressions for $D$ from Eq. (\ref{schiri1}), obtaining
\begin{eqnarray}
D= \left[  \rho_D^{\theta +1}\left(1-\nu \varrho  \right) +\chi  \right]a^{3\left( \theta +1 \right)\left(1-\nu \varrho  \right)}.\label{murano76}
\end{eqnarray}
Inserting the expression of $\chi$ derived in Eq. (\ref{murano75}) into the expression of $D$ obtained in Eq. (\ref{murano76}), we have
\begin{eqnarray}
D= a^{3\left( \theta +1 \right)\left(1-\nu \varrho  \right)}\rho_D^{\theta +1}\left(1-\nu \varrho  + 3\nu H \rho_D^{-1/2} + \omega_D   \right).\label{murano77}
\end{eqnarray}
Using in Eqs. (\ref{murano75}) and (\ref{murano77}) the expression of the EoS parameter for the R-PLECHDE model obtained in Eq. (\ref{schirinzi}), we obtain
\begin{eqnarray}
\chi_{ pl} &=& \rho_{D pl}^{\theta +1}\left[ 3\nu H \rho_{D pl}^{-1/2}-\frac{1}{3}\left(\frac{1}{3c^2 -\delta R^{\gamma /2-1}} -\frac{1+\Omega_k}{\Omega_{D pl}}\right)   \right] \nonumber \\
&=&\rho_{D pl}^{\theta +1}\left\{ 3\nu H \rho_{D pl}^{-1/2}-\frac{1}{3}\left[\frac{1}{3c^2 -\delta R^{\gamma /2-1}} -\left(1+u_{pl}\right)\right]   \right\},\label{murano78}\\
D_{ pl}&=& a^{3\left( \theta +1 \right)\left(1-\nu \varrho  \right)}\rho_{D pl}^{\theta +1} \times \nonumber   \\
&&\left[1-\nu \varrho  + 3\nu H \rho_D^{-1/2} -\frac{1}{3}\left(\frac{1}{3c^2 -\delta  R^{\gamma/2-1}}-\frac{1+\Omega_k}{\Omega_{D pl}}\right)  \right]\nonumber \\
&=& a^{3\left( \theta +1 \right)\left(1-\nu \varrho  \right)}\rho_{D pl}^{\theta +1} \times \nonumber   \\
&&\left\{1-\nu \varrho  + 3\nu H \rho_D^{-1/2} -\frac{1}{3}\left[\frac{1}{3c^2 -\delta  R^{\gamma/2-1}}-\left(1+u_{pl}\right)\right]  \right\}.\label{murano79}
\end{eqnarray}
In the limiting case of a flat Dark Dominated Universe, we obtain the following expressions for $\chi_{Dark}  $ and $D_{Dark}$
\begin{eqnarray}
\chi_{ Dark} &=&  \left[ 3  \left(\frac{6c^2}{12c^2-1}\right)^2\left(\frac{1}{t^2}\right)  \right]^{\theta +1}\times \nonumber \\
&&\left\{ \sqrt{3}\nu   +  \frac{1}{3}-\frac{1}{9c^2}    \right\},\label{murano82}\\
D_{ Dark}&=& t^{18\left( \theta +1 \right)\left(1-\nu   m_p^{-1}\right)c^2/(12c^2-1)} \left[ 3  \left(\frac{6c^2}{12c^2-1}\right)^2\left(\frac{1}{t^2}\right)  \right]^{\theta +1}\times \nonumber \\
&&\left\{ \frac{4}{3}-\nu  m_p^{-1} + \sqrt{3}\nu  -\frac{1}{9c^2}  \right\},
\label{murano83}
\end{eqnarray}
where the results of Eqs. (\ref{rhoschi}) and (\ref{LEos}) for $\rho_D$ and $\omega_D$ are utilized.\\
Using the value of $c^2$ found in the work of Gao, i.e., $c^2\approx 0.46$, we can write
\begin{eqnarray}
\chi_{ Dark} &\approx&   \left(\frac{1.119}{t^2}\right)^{\theta +1}\times \nonumber \\
&&\left(  1.732\nu  + 0.092    \right),\label{murano82}\\
D_{ Dark}&\approx&   \left(\frac{1.119}{t^2}\right)^{\theta +1} t^{1.838\left( \theta +1 \right)\left(1-\nu   m_p^{-1}\right)} \times \nonumber \\
&&\left( 1.092-\nu  m_p^{-1} + 1.732\nu    \right).
\label{murano83}
\end{eqnarray}
We must emphasize that, for the flat Dark Dominated case, we used that fact that $r_m=0$, therefore we have that $\varrho  = 1$.

Furthermore, using the general definition of EoS parameter, we can rewrite Eqs. (\ref{gcg6-1}) and (\ref{gcg7-1}) as follows.
\begin{eqnarray}
    \dot{\phi}^2  &=& \left( 1+ \omega_D \right)\rho_D, \label{gcg6mura1-1-1-1-1}\\
    V\left(  \phi \right) &=& \frac{1}{2}\left( 1 - \omega_D \right)\rho_D. \label{gcg7mura2-1-1-1-1}
\end{eqnarray}
Using in Eqs. (\ref{gcg6mura1-1-1-1-1}) and (\ref{gcg7mura2-1-1-1-1}) the expression of the EoS parameter $\omega_D$ of  the R-PLECHDE model given in Eq. (\ref{schirinzi}), we can derive the kinetic and the potential terms for the R-PLECHDE model
\begin{eqnarray}
    \dot{\phi}^2_{pl}  &=& \rho_{Dpl}\left[ 1    -\frac{1}{3} \left(  \frac{1}{3c^2 -\delta R^{\gamma /2-1}}-  \frac{1+\Omega_k}{\Omega_{Dpl}}\right)    \right] \nonumber \\
 &&\rho_{Dpl}\left\{ 1    -\frac{1}{3} \left[  \frac{1}{3c^2 -\delta R^{\gamma /2-1}}-  \left( 1+u_{pl}  \right)    \right]    \right\}, \label{gcg6mura1model1-1-1-1-1}\\
    V\left(  \phi \right)_{pl}  &=& \frac{\rho_{Dpl}}{2}\left[ 1 +\frac{1}{3} \left(  \frac{1}{3c^2 -\delta R^{\gamma /2-1}}-  \frac{1+\Omega_k}{\Omega_{Dpl}}\right) \right]    \nonumber \\
&& \frac{\rho_{Dpl}}{2}\left\{  1 +\frac{1}{3} \left[  \frac{1}{3c^2 -\delta R^{\gamma /2-1}}-  \left(  1+u_{pl} \right)  \right] \right\}. \label{gcg7mura2model1-1-1-1-1}
\end{eqnarray}
We can obtain the evolutionary form of the GCG integrating Eq. (\ref{gcg6mura1model1-1-1-1-1}) with respect to the scale factor $a\left(t\right)$
\begin{eqnarray}
	\phi\left(a\right)_{pl}  - \phi\left(a_0\right)_{pl}  &=&	 \int_{a_0}^{a}\sqrt{3\Omega_{Dpl}} \times \nonumber \\
&& \left[ 1    -\frac{1}{3} \left(  \frac{1}{3c^2 -\delta R^{\gamma /2-1}}-  \frac{1+\Omega_k}{\Omega_{Dpl}}\right)   \right]^{1/2} \frac{da}{a} \nonumber \\
&=& \int_{a_0}^{a}\sqrt{3\Omega_{Dpl}} \times \nonumber \\
&& \left\{  1   -\frac{1}{3}\left[ \frac{1}{3c^2 -\delta R^{\gamma /2-1}}- \left(1+u_{pl}\right)\right]   \right\}^{1/2} \frac{da}{a}, \label{gcg19-1-1-1-1}
\end{eqnarray}
where $\dot{\phi}=\phi' H$.

In the limiting case for a flat Dark Dominated Universe, i.e., when $\Omega_{Dpl}=\Omega_{Dlog}=1$, $\Omega_k=\Omega_m=0$ and $\delta=0$ for the R-PLECHDE model,
the scalar field and the potential of the GCG, respectively, reads
\begin{eqnarray}
\phi\left(t \right) &=& \frac{6c^2 }{\sqrt{3c^2  \left( 12c^2 -1\right)}}\ln \left( t \right),\label{murano12} \\
V\left(t\right)&=& \frac{6c^2\left( 6c^2+1\right)}{ \left(12c^2-1\right)^2}\left(\frac{1}{t^2}\right).\label{murano13gcg}
\end{eqnarray}
For $c^2\approx 0.46$, we obtain
\begin{eqnarray}
\phi\left(t \right) &\approx& 1.105\ln \left( t \right),\label{murano12} \\
V\left(t\right)&\approx& \left(\frac{0.508}{t^2}\right).\label{murano13gcg}
\end{eqnarray}
Moreover, after some algebraic calculations, we derive that the potential $V$ can be written as function of the scalar field $\phi$ as follows.
\begin{eqnarray}
V(\phi)&=&\left[\frac{6c^2(6c^2+1)}{(12c^2-1)^2}\right] e^{-\frac{\sqrt{3c^2(12c^2-1)}}{3c^2}\phi}.\label{murano14}
\end{eqnarray}
For $c^2\approx 0.46$, we have
\begin{eqnarray}
V(\phi)&\approx&0.508  e^{-1.810\phi}.\label{murano14}
\end{eqnarray}

\subsection{Dirac-Born-Infeld (DBI) Model}

We now consider the Dirac-Born-Infeld (DBI) Model. Many works with the aim of the connection between the string theory and the inflation have been recently considered. While doing so, various ideas of  string theory based on the concept of branes have proved themselves fruitful. One area which has been well explored in recent years, is inflation driven by the open string sector through dynamical Dp-branes. This is the so-called Dirac-Born-Infeld (DBI) inflation, which lies in a special class of K-inflation models. Considering the DE scalar field is a Dirac-Born-Infeld DBI scalar field, the action $ds_{bdi}$ of the field can be written as follows \cite{dbi1,dbi3,dbi4,dbi5,dbi6,dbi7,dbi8,dbi9,dbi10,dbi11,dbi12}.
\begin{eqnarray}
ds_{dbi} = \int d^4x \, a^3\left( t\right)\left[ F\left(\phi\right)\sqrt{1-\frac{\dot{\phi}^2}{F\left(\phi\right)}} + V\left(\phi\right) - F\left(\phi\right)   \right],\label{murano86}
\end{eqnarray}
where the quantity $F\left(\phi\right)$ is the tension and the quantity $V \left(\phi\right)$ represents the potential. From the action of the DBI model given in Eq. (\ref{murano86}), the corresponding pressure $p_{dbi}$
and energy density $\rho_{dbi}$ of the scalar field model can be written as follows.
\begin{eqnarray}
p_{dbi} &=&\left( \frac{\gamma-1}{\gamma}\right) F\left(\phi\right) -  V\left(\phi\right),\label{murano87}\\
\rho_{dbi} &=& \left(\gamma-1\right) F\left(\phi\right) +  V\left(\phi\right),\label{murano88}
\end{eqnarray}
where the quantity $\gamma$ is reminiscent from the usual relativistic Lorentz factor and it is given by the following relation:
\begin{eqnarray}
\gamma = \left[  1-\frac{\dot{\phi}^2}{F\left(\phi\right)}\right]^{-1/2}.\label{murano89}
\end{eqnarray}
Using the expressions of $p_{dbi}$ and $\rho_{dbi}$ given in Eqs. (\ref{murano87}) and (\ref{murano88}), we have that the EoS parameter $\omega_{dbi}$ is given by
\begin{eqnarray}
\omega_{dbi} &=& \frac{\left(\frac{\gamma-1}{\gamma}\right) F\left(\phi\right) -  V\left(\phi\right)}{\left(\gamma-1\right) F\left(\phi\right) +  V\left(\phi\right)} = \frac{\left(\gamma-1\right) F\left(\phi\right) -  \gamma V\left(\phi\right)}{\gamma\left[ \left(\gamma-1\right) F\left(\phi\right) +  V\left(\phi\right)  \right] }.\label{murano90}
\end{eqnarray}
We now want to obtain the relations for $F$, $\dot{\phi}^2$ and $V$. Adding Eqs. (\ref{murano87}) and (\ref{murano88}) and using the general definition of $\omega_D$, after some algebraic calculations, we derive that
\begin{eqnarray}
F &=& \rho_D\left( \frac{\gamma}{\gamma^2 -1}\right) \left( \omega_D + 1   \right)\label{murano91old}.
\end{eqnarray}
From the definition of $\gamma$ given in Eq. (\ref{murano89}) and using the result of Eq. (\ref{murano91old}), we obtain
\begin{eqnarray}
\dot{\phi} &=& \sqrt{\frac{ \rho_D \left( \omega_D + 1   \right) }{\gamma}}.\label{murano92old}
\end{eqnarray}
Subtracting Eqs. (\ref{murano87}) and (\ref{murano88}) and using the general definition of $\omega_D$, after some algebraic calculations, we can derive that
\begin{eqnarray}
V &=& - \left(\frac{\rho_D}{\gamma +1}\right) \left( \gamma \omega_D -1   \right).\label{murano93}
\end{eqnarray}
Therefore, we obtain the following set of expressions for $F$, $\dot{\phi}$ and $V$
\begin{eqnarray}
F &=& \rho_D\left( \frac{\gamma}{\gamma^2 -1}\right) \left( \omega_D + 1   \right),\label{murano91}\\
\dot{\phi} &=& \sqrt{\frac{ \rho_D \left( \omega_D + 1   \right) }{\gamma}},\label{murano92}\\
V &=& - \frac{\rho_D}{\gamma +1} \left( \gamma \omega_D  -1   \right).\label{murano93}
\end{eqnarray}

Using in Eqs. (\ref{murano91}), (\ref{murano92}) and (\ref{murano93}) the expression of the EoS parameter for the R-PLECHDE model given in Eq. (\ref{schirinzi}), we obtain the following expressions for $F_{pl} $, $\dot{\phi}_{pl} $ and $V_{pl} $
\begin{eqnarray}
F_{ pl} &=&\rho_{D pl} \left( \frac{\gamma}{\gamma^2 -1}\right) \left[  1 -\frac{1}{3}\left(\frac{1}{3c^2 -\delta R^{\gamma /2-1}}-\frac{1+\Omega_k}{\Omega_{D pl}}\right)  \right] \nonumber \\
 &=&\rho_{D pl} \left( \frac{\gamma}{\gamma^2 -1}\right) \left\{  1 -\frac{1}{3}\left[\frac{1}{3c^2 -\delta R^{\gamma /2-1}}-\left(1+u_{pl}\right)\right]  \right\} ,\label{murano94}\\
\dot{\phi}_{ pl} &=& \sqrt{\frac{ \rho_{D pl} \left[ 1 -\frac{1}{3}\left(\frac{1}{3c^2 -\delta R^{\gamma /2-1}}-\frac{1+\Omega_k}{\Omega_{D pl}}\right)  \right] }{\gamma}}\nonumber \\
 &=& \sqrt{\frac{ \rho_{D pl} \left\{ 1 -\frac{1}{3}\left[\frac{1}{3c^2 -\delta R^{\gamma /2-1}}-\left(1+u_{pl}\right)\right]  \right\} }{\gamma}},\label{murano95}\\
V_{ pl} &=&  \frac{\rho_{D pl}}{\gamma +1} \left[ \frac{\gamma}{3} \left(  \frac{1}{3c^2 -\delta R^{\gamma /2-1}}-\frac{1+\Omega_k}{\Omega_{D pl}} \right)  +1   \right]\nonumber \\
 &=& \frac{\rho_{D pl}}{\gamma +1} \left\{ \frac{\gamma}{3} \left[  \frac{1}{3c^2 -\delta R^{\gamma /2-1}}-\left(1+u_{pl}\right) \right]  +1   \right\}.\label{murano96}
\end{eqnarray}
In the limiting case of a flat Dark Dominated Universe, using in Eqs. (\ref{murano91}), (\ref{murano92}) and (\ref{murano93}) the expressions of $\omega_D$ and $\rho_D$ given in Eqs. (\ref{LEos}) and (\ref{rhoschi}), we find the following expressions for $F_{Dark} $, $\dot{\phi}_{Dark} $ and $V_{Dark}$
\begin{eqnarray}
F_{ Dark} &=& \left[ 3\left(\frac{6c^2}{12c^2-1}\right)^2\left(\frac{1}{t^2}\right)
  \right] \left( \frac{\gamma}{\gamma^2 -1}\right)\left( \frac{4}{3}-\frac{1}{9c^2}    \right) ,\label{murano100}\\
\dot{\phi}_{ Dark} &=& \left(\frac{6c^2}{12c^2-1}\right) \sqrt{\frac{4-\frac{1}{3c^2}}{\gamma}}\left(\frac{1}{t}\right) \nonumber \\
&=& \sqrt{\frac{12c^2}{\left( 12c^2-1\right)\gamma}}\left( \frac{1}{t}  \right),\label{murano101}\\
V_{ Dark} &=& - \frac{3\left(\frac{6c^2}{12c^2-1}\right)^2}{\gamma +1} \left[ \gamma \left( \frac{1}{3}-\frac{1}{9c^2}  \right)  -1   \right]\left(\frac{1}{t^2}\right).\label{murano102}
\end{eqnarray}
For $c^2\approx 0.46$, we have
\begin{eqnarray}
F_{ Dark} &\approx&  \left(\frac{1.222}{t^2}\right)
  \left( \frac{\gamma}{\gamma^2 -1}\right) ,\label{murano100}\\
\dot{\phi}_{ Dark} &\approx& \sqrt{\frac{1}{\gamma}}\left(\frac{1.106 }{t}\right) ,\label{murano101}\\
V_{ Dark} &\approx& - \frac{1.119}{\gamma +1} \left(\frac{ 0.092\gamma  -1   }{t^2}\right).\label{murano102}
\end{eqnarray}
We now consider the particular case corresponding to
\begin{eqnarray}
F\left( \phi  \right) = F_0 \dot{\phi}^2,\label{murano103}
\end{eqnarray}
with $F_0 >0$ being a constant parameter. In this case, we have that
\begin{eqnarray}
\gamma = \sqrt{\frac{F_0}{F_0-1}}.\label{murano104}
\end{eqnarray}
Eq. (\ref{murano104}) implies that $F_0>1$ in order to have a real value of $\gamma$. Negative values of $F_0$ also lead to positive values of $\gamma$ but they cannot be considered since we previously imposed that $F_0 >0$.

Using the result of Eq. (\ref{murano104}), we can rewrite Eqs. (\ref{murano91}), (\ref{murano92}) and (\ref{murano93}) in the following forms:
\begin{eqnarray}
F &=& \rho_D\sqrt{F_0\left( F_0-1  \right)} \left( \omega_D + 1   \right),\label{murano105} \\
\dot{\phi} &=&\left(\sqrt{\frac{F_0}{F_0-1}}\right)^{-1/4} \sqrt{ \rho_D \left( \omega_D + 1   \right)},\label{murano106} \\
V &=& - \frac{\rho_D }{\sqrt{\frac{F_0}{F_0-1}} + 1} \left[ \left(\sqrt{\frac{F_0}{F_0-1}}\right) \omega_D -1 \right].\label{murano107}
\end{eqnarray}
Using in Eqs. (\ref{murano105}), (\ref{murano106}) and (\ref{murano107}) the expression of the EoS parameter for the R-PLECHDE model given in Eq. (\ref{schirinzi}), we obtain the following expressions for $F_{pl} $, $\dot{\phi}_{pl} $ and $V_{pl} $
\begin{eqnarray}
F_{ pl} &=& \sqrt{F_0\left( F_0-1  \right)}\rho_{D pl} \left[ 1 -\frac{1}{3}\left(\frac{1}{3c^2 -\delta R^{\gamma /2-1}}-\frac{1+\Omega_k}{\Omega_{D pl} }\right)  \right]\nonumber \\
 &=&\sqrt{F_0\left( F_0-1  \right)}\rho_{D pl}  \left\{ 1 -\frac{1}{3}\left[\frac{1}{3c^2 -\delta R^{\gamma /2-1}}-\left(1+u_{pl}\right)\right]  \right\},\label{murano108} \\
\dot{\phi}_{ pl} &=&\left(\sqrt{\frac{F_0}{F_0-1}}\right)^{-1/4} \sqrt{ \rho_{D pl}  \left[  1 -\frac{1}{3}\left(\frac{1}{3c^2 -\delta R^{\gamma /2-1}}-\frac{1+\Omega_k}{\Omega_{D pl} }\right)  \right]}\nonumber \\
 &=&\left(\sqrt{\frac{F_0}{F_0-1}}\right)^{-1/4} \sqrt{ \rho_{D pl}  \left\{  1 -\frac{1}{3}\left[\frac{1}{3c^2 -\delta R^{\gamma /2-1}}-\left(1+u_{pl}\right)\right]  \right\}},\label{murano109} \\
V_{ pl} &=& \frac{\rho_{D pl}  }{\sqrt{\frac{F_0}{F_0-1}} + 1} \left\{ \left(\sqrt{\frac{F_0}{F_0-1}}\right) \left[ \frac{1}{3}\left(\frac{1}{3c^2 -\delta R^{\gamma /2-1}}-\frac{1+\Omega_k}{\Omega_{D pl} }\right)\right]+1 \right\}\nonumber \\
 &=&  \frac{\rho_{D pl}  }{\sqrt{\frac{F_0}{F_0-1}} + 1} \left\{ \left(\sqrt{\frac{F_0}{F_0-1}}\right) \left[ \frac{1}{3}\left(\frac{1}{3c^2 -\delta R^{\gamma /2-1}}-\left(1+u_{pl}\right)\right)\right]+1 \right\}.\label{murano110}
\end{eqnarray}

In the limiting case of a flat Dark Dominated Universe, using the expression of $\rho_D$ and $\omega_D$ given in Eqs. (\ref{rhoschi}) and (\ref{LEos}),  we find the following expressions for $F_{Dark} $, $\dot{\phi}_{Dark} $ and $V_{Dark}$
\begin{eqnarray}
F_{ Dark} &=& \sqrt{F_0\left( F_0-1  \right)}\left[  3\left(\frac{6c^2}{12c^2-1}\right)^2\left(\frac{1}{t^2}\right)  \right] \left( \frac{4}{3}-\frac{1}{9c^2}   \right),\label{murano114} \\
\dot{\phi}_{ Dark} &=&  \left(\sqrt{\frac{F_0}{F_0-1}}\right)^{-1/4} \left(\frac{c^2}{12c^2-1}\right)^{1/2}  \left(\frac{2\sqrt{3}}{t}\right),\label{murano115} \\
V_{ Dark} &=& - \frac{3\left(\frac{6c^2}{12c^2-1}\right)^2}{\sqrt{\frac{F_0}{F_0-1}} + 1} \left[ \left(\sqrt{\frac{F_0}{F_0-1}}\right) \left( \frac{1}{3}-\frac{1}{9c^2}  \right) -1\right]\left(\frac{1}{t^2}\right).\label{murano116}
\end{eqnarray}
From Eq. (\ref{murano115}), we can easily obtain the following relation for $\phi$
\begin{eqnarray}
\phi_{ Dark} =\left(\sqrt{\frac{F_0}{F_0-1}}\right)^{-1/4}\left(\frac{c^2}{12c^2-1}\right)^{1/2}2\sqrt{3} \ln t
\end{eqnarray}
For $c^2\approx 0.46$, we obtain
\begin{eqnarray}
F_{ Dark} &\approx& 1.092\sqrt{F_0\left( F_0-1  \right)}\left[  \left(\frac{1.119}{t^2}\right)  \right] ,\label{murano114} \\
\phi_{ Dark} &\approx& 1.105\left(\sqrt{\frac{F_0}{F_0-1}}\right)^{-1/4} \ln t \\
V_{ Dark} &\approx& - \frac{1.119}{\sqrt{\frac{F_0}{F_0-1}} + 1} \left[ 0.092\left(\sqrt{\frac{F_0}{F_0-1}}\right)  -1\right]\left(\frac{1}{t^2}\right).\label{murano116}
\end{eqnarray}

\subsection{Yang-Mills (YM) Model}
We now consider the Yang-Mills (YM) model. Recent studies suggest that Yang-Mills field \cite{ym1,ym2,ym3,ym6,ym9-1,ym9-2,ym9-3,ym9-4,ym9-5,ym9-7} can be considered as a useful candidate to describe the DE nature. There are two main reasons which can be taken into account in order to consider the YM field as source of DE. First of all, for the normal scalar field models, the connection of the field to particle physics models has not been clear until now. The second reason is that the weak energy condition can not be violated by the field. The YM field we consider has some interesting features: it is an important  cornerstone for any particle physics model with interactions mediated by gauge bosons, so it can be incorporated into a sensible unified theory of particle physics. Moreover, the EoS of matter for the effective YM Condensate (YMC) is different from that of the ordinary matter as well as the scalar fields, and the state of $-1< \omega <0$ and $\omega <-1$ can be also naturally realized.

In the effective YMC DE model, the effective Yang-Mills field Lagrangian $L_{YMC}$ is given by
\begin{eqnarray}
L_{YMC} = \frac{bF}{2}\left( \ln \left|\frac{F}{\kappa^2}\right| -1 \right),\label{murano117}
\end{eqnarray}
where the quantity $\kappa$ represents the re-normalization scale with dimension of squared mass and $F$ plays the role of the order parameter of the YMC and it is given by
\begin{eqnarray}
F= -\frac{1}{2}F_{\mu \nu}^{\alpha}F^{\alpha \mu \nu} = E^2 - B^2.\label{murano118}
\end{eqnarray}
For the  pure electric case we have, $B = 0$ which implies that $F = E^2$.\\
Furthermore, we have that $b$ is the Callan-Symanzik coefficient \cite{ym18,ym18-1} and it is given, for $SU\left(N\right)$ by the following relation:
\begin{eqnarray}
b=\frac{11N - 2N_f}{24\pi^2},\label{murano119}
\end{eqnarray}
where $N_f$ represents the number of quark flavors.

For the gauge group $SU(2)$, we have $b= 2\cdot\frac{11}{24\pi^2}$ when the fermion's contribution is neglected, and $b = 2\cdot\frac{5}{24\pi^2}$ when the number of quark flavors is taken to be
$N_f = 6$. For the case of $SU(3)$ the effective Lagrangian in Eq. (\ref{murano117}) leads to a phenomenological description of the asymptotic freedom for the quarks inside hadrons \cite{ym21,ym21-1,ym21-2}. It should be noticed that the $SU(2)$ YM field is introduced here as a model for the cosmic dark energy, it may not be directly identified as the QCD gluon fields, nor the weak-electromagnetic unification gauge fields, such as $Z^0$ and $W^{\pm}$. The YMC has an energy scale characterized by the parameter $\kappa^{1/2}\approx 10^{-3}eV$, much smaller than that of QCD and of the weak-electromagnetic unification. An explanation can be given for the form in Eq. (\ref{murano117}) as an effective Lagrangian up to 1-loop quantum correction \cite{ym21,ym21-1,ym21-2}. A classical $SU(N)$ YM field Lagrangian is given by
\begin{eqnarray}
L=\frac{1}{2g_0^2}F,\label{murano120}
\end{eqnarray}
where $g_0$ indicates the bare coupling constant. When the 1-loop quantum corrections are also included, the
bare coupling constant $g_0$ will be replaced by the running coupling $g$ as follows.
\begin{eqnarray}
g_0^2 \rightarrow g^2 = \frac{4\cdot 12\pi^2}{11N \ln \left( \frac{k}{k_0^2}  \right)} = \frac{2}{b\ln \left( \frac{k}{k_0^2}\right)},\label{murano121}
\end{eqnarray}
where $k$ represents the momentum transfer and $k_0$ indicates the energy scale. In order to have an effective theory, we can
just replace the momentum transfer $k^2$ by the field strength $F$ in the following way:
\begin{eqnarray}
\ln \left( \frac{k}{k_0^2}\right) \rightarrow 2 \ln \left| \frac{F}{\kappa^2 e} \right| = 2 \ln \left| \frac{F}{\kappa^2} -1 \right|.\label{murano122}
\end{eqnarray}
We have to notice that the result of Eq. (\ref{murano122}) yields the result of Eq. (\ref{murano117}).

Some of the interesting characteristics of this effective YMC action include the Lorentz invariance \cite{Tawfik:2012hz}, the gauge invariance, the asymptotic freedom and the correct trace anomaly \cite{Lorentz}. Having a logarithmic dependence on the field strength, the Lagrangian of the YMC model has a form similar to he Coleman-Weinberg scalar effective potential \cite{ym19} and to the Parker-Raval effective gravity Lagrangian \cite{ym20}.

We also want to emphasize that the renormalization scale $\kappa$ is the only parameter of this effective YMC model. In contrast to the scalar field DE models, this YMC Lagrangian is completely fixed by quantum corrections up to order of 1-loops, and there is no way for adjusting its functional form.

From the Lagrangian given in Eq. (\ref{murano117}), we can derive the expressions of the energy density $\rho_y$ and of the pressure $p_y$ of the YMC as follows.
\begin{eqnarray}
\rho_y &=& \frac{\epsilon E^2 }{2}+ \frac{bE^2}{2},\label{murano123}\\
p_y &=& \frac{\epsilon E^2 }{6}- \frac{bE^2}{2},\label{murano124}
\end{eqnarray}
where $\epsilon$ represents the dielectic constant of the YMC which is given by the following relation:
\begin{eqnarray}
\epsilon = 2\frac{\partial L_{eff}}{\partial F} = b \ln \left|\frac{F}{\kappa^2}\right|.\label{murano125}
\end{eqnarray}
Eqs. (\ref{murano123}) and (\ref{murano124}) can be also rewritten in the following alternative way:
\begin{eqnarray}
\rho_y &=& \frac{1}{2}b\kappa^2\left( y+1  \right)e^y, \label{murano126}\\
p_y &=& \frac{1}{6}b\kappa^2\left( y-3  \right)e^y,\label{murano127}
\end{eqnarray}
or, equivalently, as follows.
\begin{eqnarray}
\rho_y &=& \frac{1}{2}\left( y+1  \right)bE^2,\label{murano128} \\
p_y &=& \frac{1}{6}\left( y-3  \right)bE^2,\label{murano129}
\end{eqnarray}
where the parameter $y$ is defined as follows.
\begin{eqnarray}
y = \frac{\epsilon}{b} = \ln \left|\frac{F}{\kappa^2}\right| = \ln \left|\frac{E^2}{\kappa^2}\right|.\label{defiy}
\end{eqnarray}
Therefore, considering the expressions of $\rho_y$ and $p_y$ given in Eqs. (\ref{murano126}) and (\ref{murano127}) or, equivalently, in Eqs. (\ref{murano128}) and (\ref{murano129}), we obtain that the EoS parameter $\omega_y$ of the YMC model is given by
\begin{eqnarray}
\omega_y = \frac{p_y}{\rho_y} = \frac{y-3}{3\left(y+1\right)}.\label{omegay}
\end{eqnarray}
Using the definition of $y$ given in Eq. (\ref{defiy}), we can easily derive that at the critical point $\varepsilon =0$ leads to $\omega_y = -1$, then the Universe has exactly a de Sitter expansion. Near this point, if $\varepsilon <0$, we have that $\omega_y<-1$, while $\varepsilon >0$ implies $\omega_y>-1$. So, as stated before, the range of values EoS  $0>\omega_y >-1$ and $\omega<-1$  can all be naturally realized.

The expression of $\omega_y$ given in Eq. (\ref{omegay}) leads to the following expression for $y$
\begin{eqnarray}
y = - \frac{3\left(\omega_y+1\right)}{3\omega_y-1}.\label{murano130}
\end{eqnarray}
We have that,  in order to ensure that the energy density $\rho_y$ is positive in any physically viable model, the quantity $y$ should be greater than 1, which leads to $F > \kappa^2/e \approx 0.368 \kappa^2$. Before considering a particular cosmological model,  $\omega_y$  as a function of $F$ is interesting to be studied. The YMC model exhibits an EoS of radiation with $p_y = \frac{1}{2}\rho_y$ and EoS parameter $\omega_y = \frac{1}{2}$ for  large values of the dielectric, i.e., $\epsilon >>b$ (which implies $F >>\kappa^2$). On the other hand, for $\epsilon = 0$ (i.e., for $F = \kappa^2$), which is called the critical point, the YMC has an EoS of the cosmological constant, i.e., $\omega_y = -1$, with $p_y = -\rho_y$. The latter case occurs when the YMC energy density takes on the value of the critical energy density $\rho_y = \frac{1}{2}b\kappa^2$ \cite{ym1,ym3}. It is this interesting property of the EoS of YMC, going from $\omega_y = \frac{1}{3}$ at higher energies ($F >> \kappa^2$) to $w = -1$ at low energies ($F = \kappa^2$), that makes it possible for the scaling solution \cite{ym10,ym10-1} for the DE component to exist in our model. More interestingly, this transition is smooth since $\omega$ is smooth function of $y$ in the range $\left(-1, \infty \right)$. We now need to determine if the EoS parameter  $\omega_y$ can cross over $-1$. By looking at Eq. (\ref{omegay}) for $\omega_y$, we see that $\omega_y$ only depends on the value of the condensate strength $F$. In principle, $\omega_y < -1$ can be achieved as soon as $F < \kappa^2$. Moreover, with regards to the behavior of $\omega_y$ as a function of $F$, this crossing is also smooth. However, when the YMC is put into a cosmological model as DE component, together with the other components, to drive the expansion of the Universe, the value of $F$ can not be arbitrary, it comes out as a function of time $t$ and has to be determined by the dynamic evolution. Specifically, when the YMC does not decay into the matter and radiation, $\omega_y$ can only approaches to $-1$ asymptotically, but will not cross over $-1$. On the other hand, when the YMC decays into the matter and/or radiation, $\omega_y$ does cross over $-1$, and, depending on the strength of the coupling, $\omega_y$ will settle down to an asymptotic value  $-1.17$. As a merit, in this lower region of $\omega_y < -1$, all the physical quantities $\rho_y$, $p_y$ and $\omega_y$ behave smoothly, there is no finite-time singularities that are suffered by a class of scalar models.

If we now equate the EoS of the YMC $\omega_y$ with the EoS parameters of the model we are studying, we can write $y$ as follows.
\begin{eqnarray}
y = - \frac{3\left(\omega_D+1\right)}{3\omega_D-1}.\label{murano131}
\end{eqnarray}
Using in Eq. (\ref{murano131}) the expression of the EoS parameter for the R-PLECHDE model given in Eq. (\ref{schirinzi}), we obtain
\begin{eqnarray}
y_{ pl} &=& - \frac{\left[-\frac{1}{3c^2 -\delta R^{\gamma /2-1}}+\left(\frac{1+\Omega_k}{\Omega_{D pl} }\right)+3\right]}{\left[-\frac{1}{3c^2 -\delta R^{\gamma /2-1}}+\left(\frac{1+\Omega_k}{\Omega_{D pl} }\right)\right]-1}\nonumber \\
 &=& - \frac{\left[-\frac{1}{3c^2 -\delta R^{\gamma /2-1}}+\left(1+u_{pl}\right)+3\right]}{\left[-\frac{1}{3c^2 -\delta R^{\gamma /2-1}}+\left(1+u_{pl}\right)\right]-1}.\label{murano132}
\end{eqnarray}
In the limiting case of a flat Dark Dominated Universe, using the expression of $\omega_D$ given in Eq. (\ref{LEos}), we get
\begin{eqnarray}
y_{ Dark} =  12c^2-1.\label{murano134}
\end{eqnarray}
In order to have $y>1$, we derive from Eq. (\ref{murano134}) that $c^2>1/6$. The value of $c^2$ derived in the original work of RDE of Gao is $c^2 \approx 0.46$, which is in agreement with the constrain we obtained. In particular, we obtain, for $c^2 \approx 0.46$, that $y_{Dark} \approx 4.52$.

\subsection{Non-Linear Electro-Dynamics (NLED) Model}
We now consider the last model we have chosen to study, i.e., the Non-Linear Electro-Dynamics (NLED) Model. Recently, a new approach has been considered in order to avoid cosmic singularity through a nonlinear extension of the Maxwell electromagnetic theory.  Exact solutions of Einstein's field equations coupled with Non-Linear Electro-Dynamics (NLED) reveal an acceptable nonlinear effect in strong gravitational and magnetic fields. Also, the General Relativity (GR) coupled with NLED effects can explain primordial inflation.

The Lagrangian density $L_M$ for free fields in the Maxwell electrodynamics may be written as follows \cite{ele1,ele2,ele3}.
\begin{eqnarray}
L_M = - \frac{F^{\mu \nu}F_{\mu \nu}}{4\mu} , \label{murano135}
\end{eqnarray}
where $F^{\mu \nu}$ is the electromagnetic field strength tensor and $\mu$ is the magnetic permeability.
We now consider the generalization of Maxwell's electro-magnetic Lagrangian up to the second-order terms of
the fields as follows.
\begin{eqnarray}
L= -\frac{F}{4\mu_0} + \omega F^2 + \eta F^{*2},\label{murano136}
\end{eqnarray}
where $\omega$ and $\eta$ are two arbitrary constants and $F^*$ is defined as follows.
\begin{eqnarray}
F^*= F_{\mu \nu}^*F_{\mu \nu},\label{murano137}
\end{eqnarray}
where $F_{\mu \nu}^*$ is the dual of $F_{\mu \nu}$.

We now consider the particular case when the homogeneous electric field $E$ in plasma gives rise to an
electric current of charged particles and then rapidly decays. So, the squared magnetic field $B^2$ dominates over
$E^2$, i.e., $E^2 \approx 0$ and hence $F = 2B^2$. So $F$ is now only a function of the magnetic field $B$.
The pressure $p_{NLED}$ and the energy density $\rho_{NLED}$ of the Non-linear Electrodynamics Field are given, respectively, by
\begin{eqnarray}
p_{NLED} &=& \frac{B^2}{6\mu}\left(1-40\mu \omega B^2   \right), \label{murano138}\\
\rho_{NLED} &=& \frac{B^2}{2\mu}\left(1-8\mu \omega B^2   \right). \label{murano139}
\end{eqnarray}
The weak condition (given by $\rho_{NLED} > 0$) is obeyed for $B < \frac{1}{2\sqrt{2\mu \omega}}$ and the pressure $p_{NLED}$ will be negative when $B > \frac{1}{2\sqrt{10\mu \omega}}$ . The magnetic field generates DE if the strong energy condition is violated, i.e., if $\rho_B +3p_B < 0$, if $B > \frac{1}{2\sqrt{6\mu \omega}}$.

The EoS parameter $\omega_{NLED}$ of the Nonlinear Electrodynamics Field is given as follows.
\begin{eqnarray}
\omega_{NLED} = \frac{p_{NLED}}{\rho_{NLED}} = \frac{1-40\mu \omega B^2 }{3\left( 1-8\mu \omega B^2   \right)},\label{murano140}
\end{eqnarray}
which leads to the following expression for $B^2$
\begin{eqnarray}
B^2 &=&  \frac{1-3\omega_{NLED}}{8\mu \omega\left(5 -3\omega_{NLED}   \right) }.\label{murano141}
\end{eqnarray}
Making a correspondence between the EoS parameter of the NLED model with the EoS of the DE model we are studying, we obtain
\begin{eqnarray}
B^2 &=&  \frac{1-3\omega_{D}}{8\mu \omega\left(5 -3\omega_{D}   \right) }.\label{murano142}
\end{eqnarray}
Using in Eq. (\ref{murano142}) the expression of the EoS parameter for the R-PLECHDE model given in Eq. (\ref{schirinzi}), we obtain the following expressions for $B^2_{pl} $:
\begin{eqnarray}
B^2_{ pl} &=& \frac{1+\left( \frac{1}{3c^2 -\delta R^{\gamma /2-1}}-\frac{1+\Omega_k}{\Omega_{D pl} }  \right)}{8\mu \omega\left[5+ \left( \frac{1}{3c^2 -\delta R^{\gamma /2-1}}-\frac{1+\Omega_k}{\Omega_{D pl} }  \right)\right]}\nonumber \\
 &=& \frac{1+\left[ \frac{1}{3c^2 -\delta R^{\gamma /2-1}}-\left(1+u_{pl}\right)  \right]}{8\mu \omega\left\{5+ \left[ \frac{1}{3c^2 -\delta R^{\gamma /2-1}}-\left(1+u_{pl}\right)  \right]\right\}}.\label{murano143}
\end{eqnarray}
In the limiting case of a flat Dark Dominated Universe, using the expression of $\omega_D$ given in Eq. (\ref{LEos}), we get
\begin{eqnarray}
B^2_{ Dark}  = \frac{1}{8\mu \omega \left( 12c^2 +1  \right)}.\label{murano145}
\end{eqnarray}
Using in Eq. (\ref{murano145}) $c^2 \approx 0.46$, we obtain
\begin{eqnarray}
B^2_{Dark,Gao} \approx  \frac{1}{52.16 \mu \omega },\label{murano145gao}
\end{eqnarray}
which implies that $B_{Dark,Gao}  \approx \frac{1}{2\sqrt{13.04 \mu \omega} } $, i.e. a value lower than $\frac{1}{2\sqrt{2\mu \omega} } $ in order to produce DE.

\section{Conclusions}
The HDE model that interacts with DM in the non-flat FRW Universe (and has an IR cut-off equal to the Ricci scalar $R$) was examined in this study using an entropy-corrected version. An attempt to investigate the nature of DE within the context of quantum gravity is the HDE model. We took into account the logarithmic adjustment component to the HDE model's energy density. One of the most promising theories of quantum gravity, Loop Quantum Gravity (LQG), serves as the inspiration for the addition of correction terms to the energy density of HDE. We were able to determine the EoS parameter, deceleration parameter, and evolution of energy density parameter for the interacting R-ECHDE model by using the expression of this modified energy density. For both non-interacting Dark Sectors, we derived the expressions of the equation of state (EoS) parameter $\omega_D$, the deceleration parameter $q$. Moreover, we established a correspondence between the DE model considered and some scalar fields like the Generalized Chaplygin Gas (GCG), the Modified Chaplygin Gas (MCG), the Modified Variable Chaplygin Gas (MVCG), the New Modified Chaplygin Gas (NMCG), the Viscous Generalized Chaplygin Gas (VGCG), the Dirac-Born-Infeld,  the Yang-Mills and the Non-Linear Electro-Dynamics ones. We obtained the final expressions of some parameters which characterize the model we decided to study. These correspondences are essential to understand how various candidates of DE are mutually related to each other. The limiting case of a flat Dark Dominated Universe without entropy correction was studied in each scalar field.  Moreover,  we calculated the quantities we obtained for the limiting case of the flat Dark Dominated Universe, i.e. $\Omega_D =1$, $\Omega_m= \Omega_k = 0$. We also have calculated the expressions of the quantities derived in this paper for $c^2 =0.46$, as found in the work of \cite{Gao}.

\acknowledgments{ Surajit Chattopadhyay acknowledges IUCAA, Pune for hospitality during a scientific visit during December 2023-January 2024, when a part of the work was initiated.}

\end{document}